\documentclass[times,twocolumn,review]{elsarticle}
\usepackage{medima}
\usepackage{framed,multirow}

\usepackage{amssymb}
\usepackage{latexsym}

\usepackage{url}
\usepackage{xcolor}
\usepackage{lineno}
\usepackage{mathtools}
\usepackage{wrapfig}
\usepackage{makecell}
\usepackage{caption}
\usepackage{subfigure}
\usepackage{bm}
\usepackage{algorithmic}
\usepackage{graphicx}
\usepackage{textcomp}
\usepackage{color}
\usepackage{soul}
\usepackage{hyperref}
\usepackage{soul}
\definecolor{newcolor}{rgb}{.8,.349,.1}

\newcommand{\ours}{\texttt{LESS}{}}
\newcommand{\ie}[0]{{\textit{i.e.},}}
\newcommand{\eg}[0]{{\textit{e.g.},}}
\newcommand{\etc}[0]{{\textit{etc}}}
\journal{Medical Image Analysis}

\begin{document}

\verso{Beidi Zhao \textit{et~al.}}

\begin{frontmatter}

\title{
\ours: Label-efficient Multi-scale Learning for Cytological Whole Slide Image Screening}

\author[1]{Beidi \snm{Zhao}}
\author[1]{Wenlong \snm{Deng}}
\author[2]{Zi Han (Henry) \snm{Li}}
\author[2,3]{Chen \snm{Zhou}}
\author[2,3]{Zuhua \snm{Gao}}
\author[2,3]{Gang \snm{Wang}}
\author[1,4]{Xiaoxiao \snm{Li}\corref{cor1}}
\cortext[cor1]{Corresponding Author. Email: xiaoxiao.li@ece.ubc.ca}
\address[1]{Department of Electrical and Computer Engineering, The University of British Columbia, Vancouver, BC V6T 1Z4, Canada}
\address[2]{Department of Pathology, BC Cancer Agency, Vancouver, BC V5Z 4E6, Canada}
\address[3]{Laboratory Medicine, The University of British Columbia, Vancouver, BC
V6T 2B5, Canada}
\address[4]{Vector Institute, Toronto, ON M5G 1M1, Canada}


\begin{abstract}
In computational pathology, multiple instance learning (MIL) is widely used to circumvent the computational impasse in giga-pixel whole slide image (WSI) analysis. It usually consists of two stages: patch-level feature extraction and slide-level aggregation. Recently, pretrained models or self-supervised learning have been used to extract patch features, but they suffer from low effectiveness or inefficiency due to overlooking the task-specific supervision provided by slide labels.
Here we propose a weakly-supervised \textbf{L}abel-\textbf{E}fficient W\textbf{S}I \textbf{S}creening method, dubbed {\ours}, for cytological WSI analysis with only slide-level labels, which can be effectively applied to small datasets. First, we suggest using variational positive-unlabeled (VPU) learning to uncover hidden labels of both benign and malignant patches. We provide appropriate supervision by using slide-level labels to improve the learning of patch-level features. Next, we take into account the sparse and random arrangement of cells in cytological WSIs. To address this, we propose a strategy to crop patches at multiple scales and utilize a cross-attention vision transformer (CrossViT) to combine information from different scales for WSI classification. The combination of our two steps achieves task-alignment, improving effectiveness and efficiency. We validate the proposed label-efficient method on a urine cytology WSI dataset encompassing $130$ samples ($13,000$ patches) and FNAC 2019 dataset with $212$ samples ($21,200$ patches). The experiment shows that the proposed {\ours} reaches $84.79\%$, $85.43\%$, $91.79\%$ and $78.30\%$ on a urine cytology WSI dataset, and $96.88\%$, $96.86\%$, $98.95\%$, $97.06\%$ on FNAC $2019$ dataset in terms of accuracy, AUC, sensitivity and specificity. It outperforms state-of-the-art MIL methods on pathology WSIs and realizes automatic cytological WSI cancer screening. 
\end{abstract}


\end{frontmatter}


\section{Introduction}
\label{sec1}
\begin{figure*}[t!]
\centering
\includegraphics[width=\textwidth]{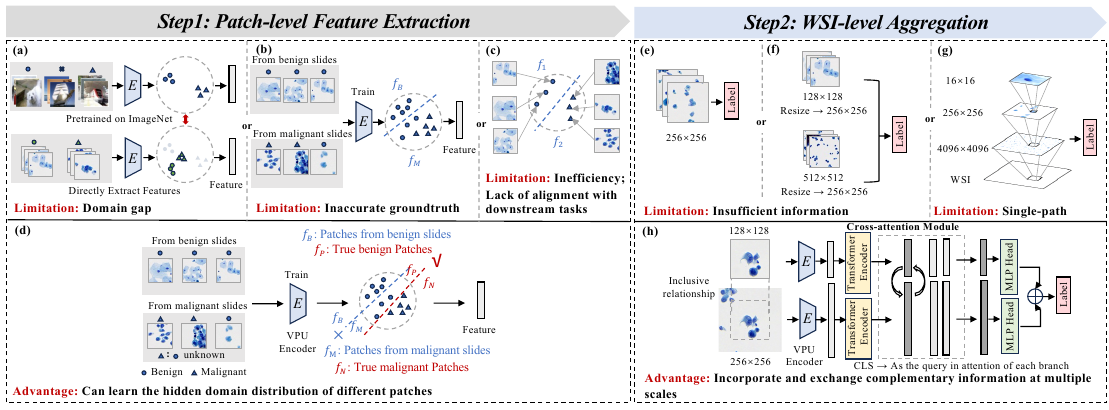}
\caption{Comparison between previous work and ours. A two-step pipline is used, including patch-level feature extraction and WSI-level aggregation. In previous work, patch features were extracted by (a) pretrained neural networks, (b) train with slide labels or (c) self-supervised contrastive learning; (d) We use a variational positive-unlabeled learning model to learn hidden true boundary between benign and malignant patches; In previous work, they use (e) single-scale fusion, (f) multi-scale with resize patches, or (g) with single-path feature extractor; (h) We simultaneously use multi-scale fusion and exchange class tokens to learn complementary information.} 
\label{fig:intro}
\end{figure*}

Cytology testing is effective, non-invasive, convenient, and inexpensive for clinical cancer screening. It falls under the field of pathology that is widely utilized for early detection of malignancies across various bodily regions, such as bladder \citep{awan2021deep,sanghvi2019performance}, cervix \citep{zhang2022whole,cheng2021robust,cao2021novel}, breast \citep{garud2017high}, lung \citep{teramoto2017automated}, and stomach \citep{li2021hybrid}. Cytology analyzes cellular morphology and compositions \citep{davey2006effect,sun2021diagnostic}. 

Specialized cytologists use ThinPrep, a commonly used liquid-based slide, to collect shed cells and examine them for abnormalities using microscopes or scanned digital whole slide images (WSIs) (with resolutions up to $10,000 \times 10,000$ pixels). Such examinations can help detect pre-cancerous changes or early-stage cancer, allowing for prompt treatment and better outcomes. However, the diagnostic evaluation of cytological specimens in clinical cytology necessitates formal training and is evaluated by international organizations \citep{jiang2022deep} and time-consuming \citep{morrison1993advantages, dey2018basic}. Therefore, there is a growing need for the development of machine learning-based autonomous screening systems that can provide accurate and efficient results.

Considering the gigapixel nature of WSIs, most existing methods adopt Multiple Instance Learning (MIL) \citep{dietterich1997solving,maron1997framework}, which consists of two stages: patch-level (or cell-level) feature extraction and slide-level aggregation.  
Given the fine-grained annotations, \citep{sanghvi2019performance,awan2021deep,li2021hybrid,zhang2022whole} have proposed leveraging supervised learning to classify cell types or detect suspicious cells. However, obtaining these fine-grained annotations can be costly, and annotations are typically disease or organ-specific, thus the algorithm cannot be generalized across diseases \citep{yu2023local}. 

This study focuses on an alternative approach to obtain the feature extractor without relying on detailed annotations, considering these limitations. Although there are increasing studies feature learning on WSI without fine-grained labels, they mainly focus on histological WSIs. 
However, these methods are not optimally designed for cytological WSIs due to their distinct properties compared with histological WSIs: (1) 
The cytological WSIs often lack detailed structural or architectural information about tissues and organs; (2) There is a scarcity of publicly available datasets, as noted in~\cite{jiang2023survey} and the available cytology datasets for specific diseases are often limited in size; (3) Cytological WSIs present more interference factors, such as extensive background noise among cells, z-stack artifacts, and cell distortion and inversion \citep{awan2021deep}.
In addition to the feature extractor, aggregation is also essential for MIL, while the current methods developed on histological WSIs fail to consider the sparsity of objects in cytological WSIs. As depicted in Figure~\ref{fig:intro}, we summarize the limitations and  provide further details in Sec.~\ref{sec:related_work}.
Therefore, it is important to propose an improved MIL method at the same time targeted to label-efficient cytological WSI analysis. 

\begin{figure*}[!t] 
\centering
\includegraphics[width=\textwidth]{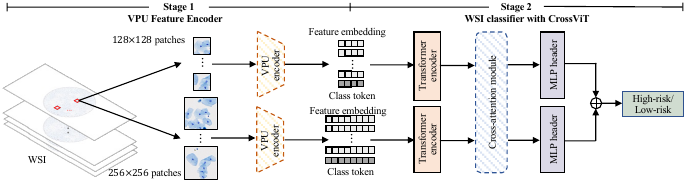}
\caption{Overview of {\ours} (a) Stage 1: WSIs are cropped to patches of two resolutions, and two VPU feature encoders are trained to get the patch feature embedding (b) Stage 2: Information at different scales is exchanged in the cross-attention module, then the model gives prediction based on the fused information. We omit the number of Transformer encoders, cross-attention modules here. Details can be found in Sec.\ref{sec:4.1.2}.} \label{fig:flowchart}
\end{figure*}

Our design is motivated by the following two insights that were previously overlooked.
Firstly, an intrinsic property of cytology cases is that malignant slides contain both benign and malignant patches, whereas benign slides only include benign patches. Secondly, information may be lost due to randomly choosing a fixed patch size and patch labels may be flipped if cropping when cropping with different resolutions.
To harness these insights, we need to reconsider two fundamental designs: \emph{(1) How to obtain reliable patch embeddings with slide-level labels and  leverage the intrinsic property from a small dataset? (2) How to use multi-scaled complementary information to enhance WSI analysis?} 

To achieve these objectives, we propose a \textbf{L}abel-\textbf{E}fficient W\textbf{S}I \textbf{S}creening method, named \ours{}, which is a two-stage, dual-path multi-scale fusion model. As illustrated in Fig.~\ref{fig:flowchart}, the first stage aims to learn the feature embedding of benign and malignant patches using only slide-labels.
We employ a weakly-supervised variational positive-unlabeled (VPU) learning model \citep{chen2020variational} to achieve this. This approach better utilizes slide-level labels, enhancing the effectiveness and task-alignment of the two-stage model. To fully harness the information from the limited dataset, learn complementary information and prevent information loss in single-scale weakly-supervised learning, the second stage employs a cross-attention-based Vision Transformer (CrossViT)  \citep{chen2021crossvit} for WSI screening. It captures and integrates information from multiple scales at the same magnification. This prevents information loss and enhances performance. The contributions of this paper are summarized as follows: 
\begin{itemize}
    \item To the best of our knowledge, we are the first to adopt a weakly-supervised variational positive-unlabelled learning method in MIL, enabling the automatic learning of the domain distributions of benign and malignant patches with only slide labels. This aligns the learning objectives of the two stages more effectively compared to other baseline patch feature learning strategies with patch annotations.
    \item We consider the unique properties of sparsity and the random distribution of cells in cytological WSIs. To address this, we propose a multi-scale learning strategy with a cross-attention module. This facilitates the exchange of multi-scale information within the same magnification but across different scales, preventing information loss and enhancing performance.
    \item We evaluate our proposed \ours{} on an in-house cytology WSI dataset with limited samples and a public high-resolution cytology image dataset.~Through extensive comparisons and ablation studies, our results demonstrate the effectiveness of \ours{} on small datasets and exhibits generalization across diverse disease classification tasks. 
\end{itemize}

\section{Related Work}\label{sec:related_work}
\subsection{Positive-Unlabeled Learning and Applications} 
Conventionally, a binary classification problem involves positive and negative labels. Positive-Unlabeled (PU) Learning aims to train a binary classifier using only a portion of the positive data and unlabeled data that contains the other positive and negative-labeled data, which includes the remaining positive and negative data. 
Most current PU learning methods  \citep{du2014analysis,hou2017generative,zhang2019positive} rely on the misclassification risk associated with supervised learning. However, such methods can be susceptible to inaccuracies in estimating class-prior probabilities. Chen et al.  \citep{chen2020variational} introduced a variational principle that enables the quantitative evaluation of modeling errors for Bayesian classifiers directly from available data. This approach results in a loss function that can be efficiently computed without the need for class prior estimation or any intermediate estimations. PU Learning has found applied to studies of medical image analysis and proved to be effective in revising incomplete annotations for chest X-rays images, histopathology images, vascular CT images, \etc . \citep{zhao2021positive,yu2022anatomy,zuluaga2011learning}. However, due to the gigabyte property of WSIs, PU learning models cannot been train on WSIs all at once. This work proposes the design of a PU learnin model at the patch-level, allowing it to learn reliable patch feature representation with the assistance of WSI labels. 

\subsection{MIL for WSI Analysis}
After \cite{campanella2019clinical} demonstrated that weakly-labeled WSI datasets outperform fine-grained annotated datasets, MIL has become the preferred method for evaluating and analyzing WSIs. However, limitation exist in both feature extraction and aggregation. 
\paragraph{Limitations of exiting patch-level (or cell-level) feature extraction methods} As shown in Fig.~\ref{fig:intro}, these methods can be further categorized into three strategies based on the learning schemes: (1) \emph{Using pretrained encoder} (Fig.~\ref{fig:intro}(a)) \citep{campanella2019clinical,lu2021data,shao2021transmil}: Extracting patch features with a frozen model pretrained on ImageNet \citep{deng2009imagenet}. (2) \emph{Supervised learning with slide-level labels} (Fig.~\ref{fig:intro}(b)): training a patch feature extractor with WSI lebels as patch labels \citep{coudray2018classification}; (3) \emph{Self-supervised contrastive learning} (Fig.~\ref{fig:intro}(c)) \citep{li2021dual,chen2022scaling}: learn the binary classifier boundary without supervision by SimCLR \citep{chen2020simple}, DINO \citep{caron2021emerging}, \etc. In the context of cytology WSIs, those methods encounter distinct challenges. Firstly, Strategy (1) experiences a decline in performance due to domain differences between pretrained images, such as natural images and cytology images \citep{chen2022self}. Secondly, Strategy (2), which involves directly assigning WSI labels to instance bags, can sometimes assign incorrect ground truths, leading to an an inaccurate distribution boundary between instances. Lastly, the success of Strategy (3) is contingent upon the availability of large datasets and massive iterations. This is problematic in the cytology domain due to the challenges of accessing extensive WSI datasets \citep{jiang2023survey}. Moreover, compared to histopathology WSIs, fewer patches of interest can be sampled from cytological WSIs due to the sparse distribution of cells, especially for benign WSIs. We aim to involve the medical background knowledge as weak supervision to learn the hidden distribution of benign and malignant patches (Fig.~\ref{fig:intro}(d)). 
\paragraph{Limitations of exiting slide-level aggregation methods} 
In addition to the feature extractor, aggregation is also essential for MIL. \cite{ilse2018attention,campanella2019clinical} pioneered the use of innovative operators, such as neural network-based permutation-invariant aggregation and mean/max-pooling, to enhance patch-level feature aggregation. This field saw further advancements with \cite{lu2021data} introducing the clustering-constrained-attention multiple-instance learning (CLAM), significantly elevating the histopathology WSI analysis. Furthermore, the Transformer architecture, initially popularized in natural language processing (NLP), has been successfully integrated into histological WSI analysis by works like TransMIL \cite{shao2021transmil}. Nonetheless,
as shown in Fig.~\ref{fig:intro}(e), these methods extract patches of a fixed size, e.g. $256 \times 256$, they may lose context information. 
Then, multi-magnification (Fig.~\ref{fig:intro}(f)) \citep{li2021dual} and pyramid-like multi-scale (Fig.~\ref{fig:intro}(g)) \citep{chen2022scaling} were proposed to learn hierarchical features by resizing patches to the same size, or sequentially train Transformer models to learn fine-grained and coarse information.They work for histopathology WSIs, since multiple magnifications can help to learn coarse-grained and fine-grained features (e.g. vessels, glands and tumors v.s. cells), but cytology is blurred for large-scale low magnification patches, uninformative for structural information, and separate single-path training process is less efficient. In contrast, we aim to employ multi-path cross-attention-based fusion to capture complementary information from multi-scaled patches (Fig.~\ref{fig:intro}(h)).

\subsection{Cytological WSI Cancer Screening}
With the prosperity of the WSI analysis in histology, MIL-based methods have also achieved satisfactory results in cytology.
\citep{awan2021deep} proposed a cell detection and classification model l based on different kinds of annotated cell, grading urine cytology WSIs based on the count of atypical and malignant cells. \citep{sanghvi2019performance} used convolutional neural networks to extract urothelial cells and cell cluster features, then they combine these features with WSI-level features to perform urine cytological WSI screening. \citep{li2021hybrid} introduced a mixed supervision learning method with sufficient WSI-level coarse annotations and a few pixel-level labels for effective gastric cytology WSI classification. 
\citep{cao2021novel} incorporated an attention module into a feature pyramid network to extract multi-scaled features of convolutional neural networks. They validated the effectiveness of the attention mechanism in tasks ranging from cell-level detection to case-level classification. \citep{zhang2022whole} proposed a whole slide cervical cancer screening model based on a graph attention network constructed by top-k and bottom-k suspicious patches.While the methods mentioned above have achieved inspiring results, they all require manually annotated cell or patch labels, which can be laborious and disease-specific, limiting their generalizability across diseases \citep{yu2023local}.

\section{Methodology}
This work aims to efficiently and effectively utilize slide-level labels to identify high-risk cancer WSIs by incorporating multi-scale information without relying on cell or patch-level annotations. In this section, we will describe the overall pipeline and introduce the key technical contributions and the rationale behind the design.

\subsection{Overview of the Proposed Pipeline}
As shown in Fig.~\ref{fig:flowchart}, the overall pipeline of our proposed method \ours{}
contains two stages, including 1) patch embedding learning and 2) multi-scale aggregation. 

\begin{figure*}[!t]
\centering
\includegraphics[width=\textwidth]{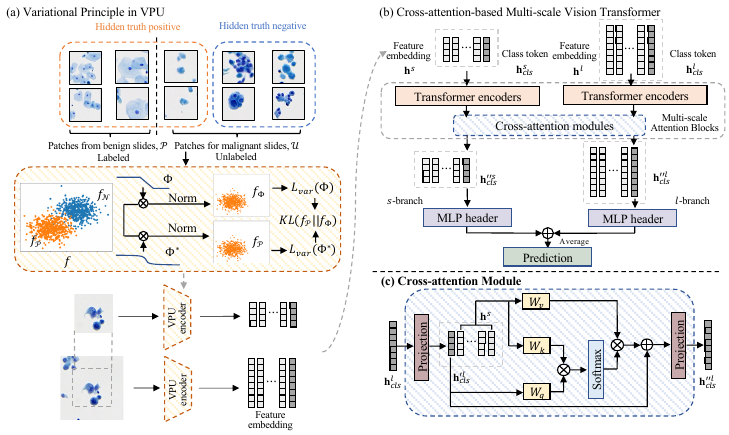}
\caption{Illustration of the variational principle used in VPU and the CrossViT structure with the cross-attention module:
(a) The VPU encoder aims to learn a Bayesian classifier from the dataset with only partial positive samples labeled.
(b) Features of multi-scale patches obtained by VPU are passed to CrossViT, providing WSI-level predictions with multi-scale information.
(c) Half of the inner structure of the cross-attention module: The class token of one branch serves as the query and interacts with information embedded in tokens from the other branch.} 
\label{fig:illustration}
\end{figure*}

We begin with cropping the WSI into patches following previous work  \citep{awan2021deep,sanghvi2019performance,li2021hybrid,cao2021novel,zhang2022whole} in the first stage.  This stage focuses on learning effective feature embeddings of patches. These patch embeddings are integrated into the fusion stage to make predictions over WSIs. 
In contrast to previous studies that: 1) rely on fine-grained labels, which can be costly; 2) employ pretrained models or assign slide-level labels as patch-level labels, which eliminates the need for fine-grained labels but may sacrifice performance; and 3) train self-supervised learning model using large amounts of data, our objective is to eliminate the need for additional labeling while simultaneously learning reliable patch features and achieving high WSI classification performance from a small dataset. However, this endeavor presents a substantial challenge: extracting fine-grained patch representations from relatively coarse label information. To overcome this challenge, we employ a weakly-supervised-learning strategy that solely leverages WSI-level labels. This is facilitated by Variational Positive-Unlabeled (VPU) learning, a mechanism capable of autonomously refining optimal labels to supervise representation learning. We will first introduce the critical novel VPU-based patch feature encoder to efficiently learn feature representation of benign and malignant patches using only WSI-level labels (Sec.~\ref{sec:3.2.1}) and then introduce the variational principle and MixUp-based consistency regularization in Sec.~\ref{sec:3.2.2}.  

Despite learning effective features, the gigabyte-sized nature of WSIs still requires a model capable of integrating this extensive information. The challenge is to design a more effective fusion that leverages as much information as possible with limited amount of data. Therefore, we introduce a multi-scale fusion method based on CrossViT \citep{chen2021crossvit} to incorporate and exchange multi-scale information in Sec.~\ref{sec:3.3}. We will begin with multi-scale feature learning in Sec.~\ref{sec:3.3.1}, then introduce the structure of multi-scale cross-attention blocks in Sec.~\ref{sec:3.3.2}. 

\subsection{Patch Feature Extraction} \label{sec:3.2}
\subsubsection{Problem Formulation}\label{sec:3.2.1}
We denote patches that contain only benign cells as positive data, and those contain suspicious cells as negative data. 
Based on the fact that {\em both positive and negative patches exist in malignant WSIs, whereas benign WSIs only contain positive patches}, we introduce positive-unlabeled (PU)~ \citep{chen2020variational} learning to maximize the utilization of slide-level annotation. PU learning is a binary classification problem with input from a labeled positive dataset $\mathcal{P} = \{\bm{x}_1,\ldots,\bm{x}_M\}$ (patches from benign WSIs) and an unlabeled dataset $\mathcal{U} = \{\bm{x}_{M+1},\ldots,\bm{x}_{M+N}\}$ (patches from malignant WSIs) that contains true positive and negative samples. Assuming that labeled and unlabeled data are independently drawn as
\begin{equation}
\mathcal{P}=\left\{\bm{x}_i\right\}_{i=1}^M \stackrel{\text { i.i.d }}{\sim} f_\mathcal{P}, \quad \mathcal{U}=\left\{\bm{x}_i\right\}_{i=M+1}^{M+N} \stackrel{\text { i.i.d }}{\sim} f,
\end{equation} 
where $f_\mathcal{P}$ is the distribution of positive data, $f$ is the distribution of unlabeled data. There is a set $\mathcal{X} \subset \mathbb{R}^d$ that satisfies $\int_{\mathcal{X}} f_\mathcal{P}(\bm{x}) \mathrm{d} \bm{x}>0$ and 
\begin{equation}
\Phi^*(\bm{x})=1, \quad \forall \bm{x} \in \mathcal{X}.
\end{equation}
The goal is to approximate an ideal Bayesian classifier 
\begin{equation}
\Phi^*(\bm{x}) \triangleq \mathbb{P}(y=+1 \mid \bm{x})
\label{eq:Bayesian}
\end{equation}
that can accurately predict true positive and negative  input labels from $\mathcal{P}$ and $\mathcal{U}$. 
In our settings, the positive set $\mathcal{P}$ consists of image patches from benign WSIs, while the unlabeled set $\mathcal{U}$ comprises patches from malignant WSIs. The PU learning method then trains a binary classifier on these sets to enable accurate prediction on class labels of patches in malignant WSIs. Here, we regard $\mathcal{X} = \{\bm{x}_1,\ldots,\bm{x}_{M+N}\}$ as the patches extracted from all WSIs, we aim to use weakly-supervised learning to learn the boundary between benign and malignant patches with only a portion of labeled benign patches. 
\paragraph{Challenges of Applying PU learning in Our Problem}
Most PU learning methods  \citep{du2014analysis,kiryo2017positive,zhang2019positive,hou2017generative} require class prior estimation, which involves estimating  $\pi_\mathcal{P} = \mathbb{P} (y=+1)$, \ie{} the proportion of benign patches in all patches from malignant slides in our application. However, obtaining such information is unfeasible without fine-grained patch-level annotations. 

\subsubsection{Patch Feature Encoding with Variational PU (VPU) Learning}\label{sec:3.2.2} 
Given the fact that we do not have class prior for patch labels, we propose to use a variational positive-unlabeled (VPU) learning model  \citep{chen2020variational} as the feature encoder to learn patch-level features. This approach allows us to effectively build a feature encoder for distinguishing benign and malignant patches with only WSI-level labels.
The illustration of the variational principle is shown in Fig.~\ref{fig:illustration}(a). With the learned classifier $\Phi$ approximated to the ideal Bayesian classifier $\Phi^*$, the variational principle is driven by the approximation that
\begin{equation}
\begin{aligned}
f_\mathcal{P}(\bm{x})&=\frac{\mathbb{P}(y=+1 \mid \bm{x}) \mathbb{P}(\bm{x})}{\int \mathbb{P}(y=+1 \mid\bm{x}) \mathbb{P}(\bm{x}) \mathrm{d} \bm{x}}\\
& \approx \frac{\Phi(\bm{x}) f(\bm{x})}{\mathbb{E}_f[\Phi(\bm{x})]} \triangleq f_{\Phi}(\bm{x}) 
\end{aligned}
\end{equation}
Thus, we can minimize the KL divergence of learned positive distribution $f_\mathcal{P}$ and $f_\Phi$. The KL divergence can be written as
\begin{equation}
    KL(f_\mathcal{P}||f_{\Phi}) = L_{var}(\Phi) - L_{var}(\Phi^*),
\end{equation}
where $L_{var}(\Phi)$ is the variational loss of the learned classifier, and $L_{var}(\Phi^*)$ is the variational loss of the ideal classifier. Since $L_{var}(\Phi^*)$ is invariant, and KL-divergence is always non-negative, $L_{var}(\Phi)$  provides a variational upper bound of $L_{var}(\Phi^*)$~\citep{chen2020variational}. To minimize the KL divergence, we only need to minimize 
\begin{equation}
    L_{var}(\Phi) \triangleq \log \mathbb{E}_f[\Phi(\bm{x})] - \mathbb{E}_{f_\mathcal{P}}[\log\Phi(\bm{x})], \label{loss:v}
\end{equation}
where $f$ is the distribution of unlabeled data. As a result, The variational principle enables PU learning to learn the binary classifier without the class prior.
We utilize MixUp \citep{zhang2017mixup} to avoid overfitting caused by the deep neural network,
\begin{equation*}
\begin{aligned}
\tilde{\bm{x}}&=\gamma\cdot \bm{x}^{\mathcal{P}}+(1-\gamma)\cdot \bm{x}^{\mathcal{U}},\\
\tilde{\Phi}&=\gamma\cdot 1+(1-\gamma) \cdot\Phi\left(\bm{x}^{\mathcal{U}}\right).
\end{aligned}
\end{equation*}
where $\gamma$ can be sampled from Beta distribution, $\tilde{\bm{x}}$ is a mixed sample generated by randomly selected sample $\bm{x}^{\mathcal{P}}$ from the positive set and $\bm{x}^{\mathcal{U}}$ from the unlabeled set, $\tilde{\Phi}$ is the guessed probability $\mathbb{P}(y=+1 \mid \bm{x}=\tilde{\bm{x}})$ generated from the linear interpolation of the true label and the prediction of model $\Phi$. To this end, we introduce the additional regularization term following \cite{chen2020variational}
\begin{equation}
\mathcal{L}_{reg}(\Phi)=\mathbb{E}_{\tilde{\Phi}, \tilde{\bm{x}}}\left[\left(\log \tilde{\Phi}-\log \Phi\left(\tilde{\bm{x}}\right)\right)^2\right],
\end{equation} 
It is performed between labeled and unlabeled samples, adding smoothness to the model. 
Thus, the final objective function of the first stage is as follows
\begin{equation}
\min _{\Phi} \mathcal{L}(\Phi)=\mathcal{L}_{var}(\Phi)+\lambda \mathcal{L}_{reg}(\Phi), \label{eq:8}
\end{equation}
where $\lambda \in [0,1]$ is a parameter of the regularization term. Although VPU method was first proposed in \cite{chen2020variational} for low-resolution natural image classification (\eg UCI dataset \citep{dua2017uci}, Cifar-10 \citep{krizhevsky2009learning}, FasionMNIST \citep{xiao2017fashion}, STL-10 \citep{coates2011analysis}), to the best of our knowledge, we are the first to adapt this strategy to the giga-pixel WSI analysis by leveraging the observation described in \ref{sec:3.2.1} and demonstrating its effectiveness (as shown in Table.~\ref{tab:02-SOTA}).

\subsection{Slide-level Aggregation with Multi-scale Features} \label{sec:3.3}
\subsubsection{Multi-scale Feature Learning}\label{sec:3.3.1}
Traditional WSI analysis methods tend to use a fixed patch size (\eg $128 \times 128$), which can introduce problems for VPU when dealing with randomly distributed cells in cytology WSIs. For instance, if we crop a $256 \times 256$ patch that contains suspicious cells surrounding around a positive (benign) $128 \times 128$ patch, intuitively, the label will be flipped to negative (malignant) in VPU feature extraction (see Sec.~\ref{sec:3.2.1}). To overcome the conflict in embedding patch contents, we aim to develop a multi-scale aggregation method to capture complementary information from individual and surrounding contextual cells. Instead of resizing the patches to different magnifications used in previous histopathology WSI analysis methods \citep{chen2022scaling,li2021dual}, we recognize that resizing can make cytology patches appear blurred due to the sparse nature of objects in cytology WSIs. To address this, we propose to crop aligned patches, enlarging $128\times128$ patches to $256\times256$ and with the center of $128\times128$ patches.
Additionally, we harness the power of Vision Transformer (ViT), which has demonstrated superior capabilities in computer vision \citep{dosovitskiy2020image}. Specifically, we use features from before the last layer of VPU models as input tokens for two branches, rather than using feature embedding layers of ViT. Self-attention is applied to each branch, followed by cross-attention across branches to form the CrossViT structure \citep{chen2021crossvit} that can incorporate and exchange information between central and contextual cells as detailed in Sec.~\ref{sec:3.3.2}.

\subsubsection{Multi-Scale Aggregation}\label{sec:3.3.2}
\begin{figure}[t]
\centering
\includegraphics[width=0.9\columnwidth]{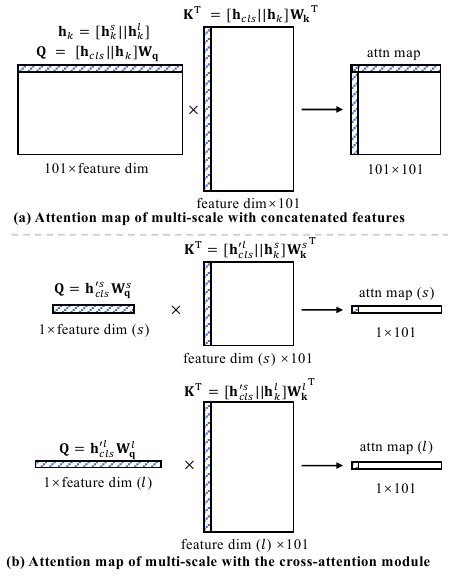}
\caption{Comparison of attention maps of concatenated features with the self-attention module and dual-path features with the cross-attention module.}\label{fig:attn}
\label{fig:dataset}
\end{figure}

Our multi-scale aggregation model (shown in Fig.~\ref{fig:illustration}(b)) consists of two parts: (1) Multi-scale attention blocks (2) MLP header for classification. The multi-scale attention blocks can be further divided into: (1a) transformer encoder blocks that include self-attention modules; (1b) cross-attention modules. The following section will provide a detailed introduction the model.
\paragraph{Transformer Encoder}
Denote the patch features encoded by VPU as $\mathbf{h}$, similar to the original Transformer used in Natural Language Processing  \citep{devlin2018bert}, a learnable class token (CLS) is added to the sequence (Eq.~\eqref{eq:ipt_patch}). Then, all tokens are fed into the transformer encoder blocks with multi-head self-attention (MSA) modules. For each self-attention module, it is combined with a two-layer multilayer perceptron (MLP) constructed feed-forward network (FFN) with GELU  \citep{hendrycks2016gaussian} after the first linear layer and dropout (Eq.~\eqref{eq:msa},\eqref{eq:mlp}). Between each transformer block, layer normalization is applied to reduce the effect of covariate shifts in the model. 
Suppose the input of ViT is $\mathbf{h}_0$, the embedding of $\mathbf{h}_0$ after the $k$-th block is
\begin{align}
\mathbf{h}_0 & =\left[\mathbf{h}_{c l s} \| \mathbf{h}\right]\label{eq:ipt_patch},\\
\mathbf{h}^{\prime}_k & =\mathbf{h}_{k-1}+\operatorname{MSA}\left(\operatorname{LN}\left(\mathbf{h}_{k-1}\right)\right), \quad &k=1 \ldots K, \label{eq:msa}\\
\mathbf{h}_k & 
=\mathbf{h}^{\prime}_{k} + \operatorname{MLP}\left(\operatorname{LN}\left(\mathbf{h}^{\prime}_{k}\right)\right), \quad &k=1 \ldots K, \label{eq:mlp}
\end{align}
where $\mathbf{h}_{c l s} \in \mathbb{R}^{1 \times C}$ is the class token, $\mathbf{h} \in \mathbb{R}^{N \times C}$ is the patch token, and $\mathbf{z}$ is the output. $N$ is the number of patch tokens and $C$ is the number of dimension of the embedding. 
\paragraph{Cross-attention Module}\label{sec:cvit2} 
Suppose the features output by the transformer encoder in each branch are $\mathbf{h}^s_k$ and $\mathbf{h}^l_k$. We then feed them into the cross-attention modules.
Fig.~\ref{fig:illustration}(c) shows the $l$-branch, which exchanges information between the large-scale class token $\mathbf{h}_{cls}^l \in \mathbb{R}^{d_{l}}$ with $d_l$ dimension and embedding tokens $\mathbf{h}^s_k$ from the $s$-branch. The projection layer $f^{l}(\cdot)$, including a LayerNorm  \citep{ba2016layer} followed by GELU  \citep{hendrycks2016gaussian} and a linear layer, is to make $\mathbf{h}_{cls}^l$ align with $\mathbf{h}^s$ in dimension 
\begin{equation}\label{eq:13}
    \mathbf{h}_{cls}^{\prime l} = f^{l}({\mathbf{h}}_{cls}^{l}) \in \mathbb{R}^{d_s}.
\end{equation} Then, $\mathbf{h}_{cls}^{\prime l}$ is concatenated with $\mathbf{h}^s$. The module conducts cross-attention (CA) between $\mathbf{h}_{cls}^{\prime l}$ and $\mathbf{h}^s$ as
\begin{align}
\mathbf{q}&=\mathbf{h}_{cls}^{\prime l} \mathbf{W}_{\mathbf{q}}, \quad \mathbf{k}=\left[\mathbf{h}_{cls}^{\prime l} || \mathbf{h}^s_k\right]\mathbf{W}_{\mathbf{k}}, \quad \mathbf{v}=\left[\mathbf{h}_{cls}^{\prime l} || \mathbf{h}^s_k\right]\mathbf{W}_{\mathbf{v}},\label{eq:14}\\
&\mathbf{A}^l = \text{Attention}(\mathbf{q},\mathbf{v},\mathbf{k}) = \text{softmax}(\mathbf{q}\mathbf{k}^\top/\sqrt{d_s/n})\mathbf{v},\label{eq:15}
\end{align}
where $\mathbf{q},\mathbf{k},\mathbf{v} \in \mathbb{R}^{\sqrt{d_s/n}}$ are query, key and value, $d_s$ is the feature embedding dimension of the $s$-branch, $n$ is the number of heads, and $\mathbf{A}^l$ is the attention map of the $l$-branch. In CA, only $\mathbf{h}_{cls}^{\prime l}$ is used as the query, as it contains information from the $l$-branch and achieves linear computational complexity. Furthermore, the feed-forward network (FFN) in the original ViT  \citep{dosovitskiy2020image} is replaced by a residual connection \citep{he2016deep} to reduce the number of parameters and preserve information from $\mathbf{h}_{cls}^l$. 
Fig.~\ref{fig:attn} shows the comparison to multi-scale concatenated feature input \citep{li2021dual} followed by self-attention. Cross-attention reduce the computational complexity to linear, and exchanging class tokens prevents information of different scales be intermingled after MLP layers. Compared to pyramid-like attention \citep{chen2022scaling}, cross-attention does not aim to get hierarchical (coarse and fine-grained) features, but rather to capture complementary features from different scales of under the same magnification. 

To this end, we get the class token $\mathbf{h}_{cls}^{\prime\prime l}$ after acquiring the information from the small-scale branch, namely
\begin{equation}
    \mathbf{h}_{cls}^{\prime\prime l} = g^{l}(f^{l}({\mathbf{h}}_{cls}^{l})+ \operatorname{MCA}\left(\mathrm { LN } \left(\left[f^l\left(\mathbf{h}_{c l s}^l\right) \| \mathbf{h}^s_k\right]\right)\right), \label{eq:16}
\end{equation}
where $\mathbf{h}_{cls}^{\prime\prime l} \in \mathbb{R}^{d_l}$,
which is then passed through the back-projection layer $g^{l}(\cdot)$ to recover to its original dimensionality and added back to $l$-branch embedding tokens. Then, the output of the cross-attention module can be written as
\begin{equation}
\mathbf{h}^l_{k+m}=\left[g^l\left(\mathbf{h}_{c l s}^{\prime \prime l}\right) \| \mathbf{h}^l_k\right].\label{eq:17}
\end{equation}
where $m$ is the number of cross-attention module blocks. The $s$-branch follows a symmetric process by exchanging notations of $l$ and $s$ in \eqref{eq:13}-\eqref{eq:17}. Transformer encoders and cross-attention modules together form the multi-scale attention block. The first part of aggregation model is consist of several stacked multi-scale attention blocks, followed by the second part of MLP headers to get the prediction.

\noindent\textbf{Summary of the Pipeline.} To sum up, the whole pipeline contains two modules, one is the VPU feature encoder, the other is cross-attention-based multi-scale ViT. The VPU model takes patches and slide-level label as input and generates patch feature embedding, which is passed to the multi-scale ViT as input and get predictions of WSIs. 

\section{Experiment}
\subsection{Experimental Setup}
\subsubsection{Dataset}
We conduct experiments on two cytological image datasets: a urine cytological WSI dataset collected by BC Cancer Agency and a public high-resolution breast image dataset. Examples of datasets used in this paper are shown in Fig.~\ref{fig:dataset}.
\begin{figure}[t]
\centering
\includegraphics[width=0.8\columnwidth]{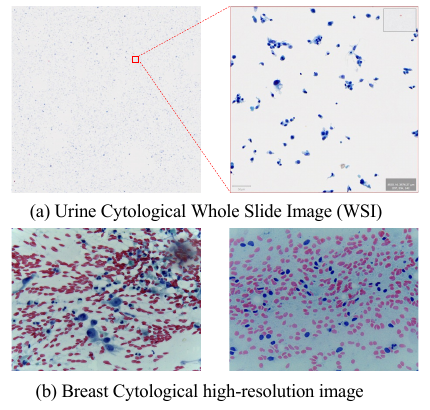}
\caption{Examples of the urine cytological WSI and high-resolution breast cytological images.}
\label{fig:dataset}
\end{figure}

\noindent\textbf{Urine Cytological WSI Dataset.} We first evaluate \ours{} on a urine cytological WSI dataset that contains $130$ urine cytology Thinprep pad specimens 
scanned on a MoticEasyScan Infinity instrument at $40\times$ magnification. All samples have been screened and diagnosed by experienced cytotechnologists, further confirmed by expert pathologists and approved by the ethics committee. 
The average size is around $60,000 \times 60,000$. The dataset contains high-risk (with subtypes of suspicious ($19$) and malignant ($41$)) and low-risk (with subtypes of benign ($42$), atypical ($28$)) samples. 

\noindent\textbf{FNAC 2019  \citep{saikia2019comparative}.} To validate the generalizability of cancer screening, we present experimental results apart from the urine cytological WSI dataset. Given the scarcity of publicly available WSI datasets, we apply our method to a high-resolution cytological image dataset FNAC 2019 \citep{saikia2019comparative} with $2,048\times 1,536$ breast cytological images. 
The dataset consists of $212$ fine-needle aspiration breast cell inspection images categorized benign ($99$) or malignant ($113$). Since our method is patch-based, they can be used comparable to WSIs. 

\subsubsection{Implement Details}\label{sec:4.1.2}
\begin{figure*}[t]
\centering
\includegraphics[width=\textwidth]{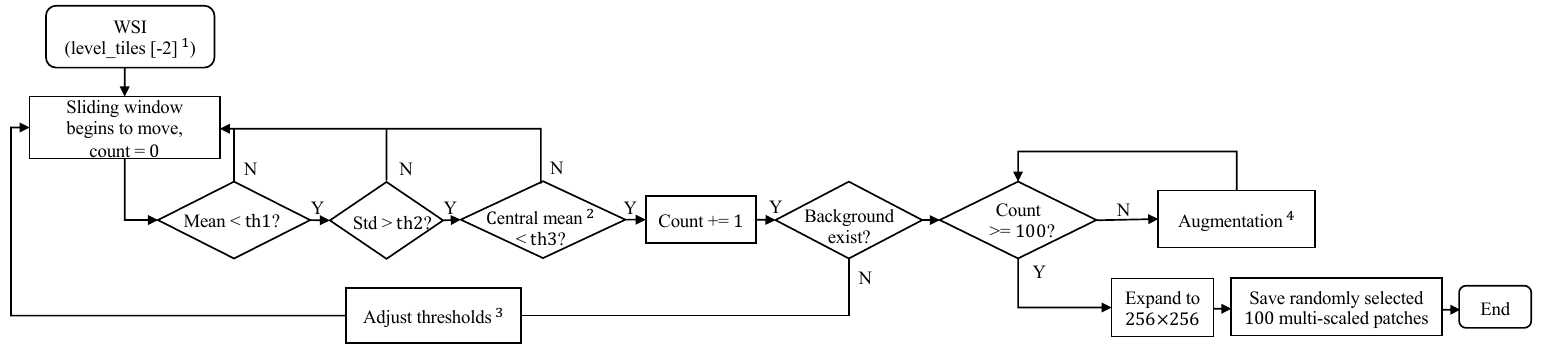}
\caption{Flowchart of WSI preprocessing. $^1$ The level\_tiles are generated using DeepZoomGenerator from the Python library Openslide; The $^2$ denotes the central ½ area of the patch; $^3$ Thresholds should be determined based on the specific dataset. In our setting, the initial thresholds are th1 = $236$, th2 = $10$, th3 = $236$. Th1 is used to filter out the background, th2 is employed to exclude patches with large blank spaces and blurriness, and th3 ensuresthat cells are not only present at the border; $^4$ Data augmentation techniques utilized in our study include rotation, left-right flipping, top-bottom flipping and zooming. 
}
\label{fig:wsi_flowchart}
\end{figure*}

\paragraph{Preprocessing} We follow a similar preprocessing to the existing MIL methods~\citep{awan2021deep,sanghvi2019performance,li2021hybrid,cao2021novel,zhang2022whole}. This involves using sliding windows with $50\%$ overlap to crop $128\times 128$ patches. Additionlly. we employ simple filters based on RGB pixel mean and standard deviation to remove background and irrelevant patches as shown in Fig.~\ref{fig:wsi_flowchart}. 
Then, we randomly select and retain $100$ patches from each WSI, with two sizes: $256\times 256$ and $128 \times 128$. The $256 \times 256$ patches are derived by expanding the central $128\times 128$ patches. In cases where there are fewer than $100$ patches available, we utilize image augmentation techniques to augment the dataset. The backbone of VPU feature encoder is a $7$-layer CNN, with four convolutional layers and three linear layers. ReLU activation is applied after each convolutional layer, and the final activation layer is LogSoftmax to get the predicted patch label. Feature embedding is extracted prior to the final linear layer. For the multi-scale aggregation model, the input comprises tokens with dimensions of $384$ from $128\times 128$ patches and $768$ from $256\times 256$ patches. The number of multi-scale attention blocks is $3$, with $4$ and $1$ transformer encoders for small and large braches and $1$ cross-attention module inside. All experiments are implemented using PyTorch on the NVIDIA $3090$Ti GPU.

\begin{table}[h]
\centering
\caption{Hyperparameter setting of models. MILs contain mean/max-pooling, HIPT, DTFT-MIL, ABMIL and CLAM-SB (learning rates are the same as original papers). Model structures and hyperparameters are the same as those in the original papers if not listed. 
}\label{tab:01-hyperparameter}
\subtable[Hyperparameters of the first stage.]{
\resizebox{6cm}{!}{%
\begin{tabular}{c|c|c|c}
\hline\hline
\multirow{2}{*}{\begin{tabular}[c]{@{}c@{}}Hyperparameter\\ Setting\end{tabular}} &
 \multicolumn{3}{c}{Stage $1$}  
  \\ \cline{2-4} 
&VPU/PN &SimCLR&DINO \\ \hline \hline
Batch size &$100$ & $256$&$64$\\ \hline
Epochs &$10$ &$50$&$50$ \\ \hline
Optimizer &Adam &Adam &Adamw  \\ \hline
\begin{tabular}[c]{@{}c@{}}Learning rate\\ (initial)\end{tabular} &$1e-4$ &$3e-4$&$5e-4$\\ \hline
LR scheduler &\begin{tabular}[c]{@{}c@{}}Step\\ ($0.5$ every $2$ epochs)\end{tabular} & Cosine& Cosine \\ \hline\hline
\end{tabular}%
}}
\subtable[Hyperparameters of the second stage.]{
\resizebox{8.8cm}{!}{
\begin{tabular}{c|c|c|c|c|c|c}
\hline\hline
\multirow{2}{*}{\begin{tabular}[c]{@{}c@{}}Hyperparameter\\ Setting\end{tabular}} &
\multicolumn{6}{c}{Stage $2$}  \\ \cline{2-7} 
& CrossViT (VPU) &CrossViT (SSL) &ViT&DSMIL&GNN-MIL&MILs\\ \hline \hline
Epochs &$40$&$200$ &$40$&$100$&$100$&$40$ \\ \hline
Optimizer &Adam &Adam &Adam&Adam &Adam &Adam \\ \hline
\begin{tabular}[c]{@{}c@{}}Learning rate\\(initial)\end{tabular}&
$1e-6$&$1e-6$&$1e-6$&$5e-3$&$4e-4$&-\\ \hline
LR scheduler& Cosine &Cosine & Cosine & Cosine & Cosine & Cosine \\ \hline
Weight decay &$5e-2$&$5e-2$&$5e-2$&$5e-3$&$0$&$1e-5$\\ \hline\hline
\end{tabular}%
}}
\end{table}

\paragraph{Data splitting and training setup} We performed a random split for training and testing five times to repeat the experiment. Benign and malignant WSIs are divided into $70\%$ training data and $30\%$ testing data to train the first stage model. Since the dataset is small, after splitting the benign and malignant WSIs, there are only $26$ WSIs left for testing, which hinders the presentation of the statistical significance of the performance results. Thus, we further involve subtypes atypical to and suspicious WSIs to the testing set to demonstrate the performance of our model more effectively. Following \citep{awan2021deep,sanghvi2019performance}, we categorize benign and atypical as low-risk and suspicious and malignant as high-risk. For our method, we train the first-stage VPU model for $10$ epochs with a batch size of $100$ and the second stage for $40$ epochs uisng five random seeds for each split. 
Hyperparameters of all models are shown in Table.~\ref{tab:01-hyperparameter}. 
 
\label{sec:performance}
\subsection{Comparison with Baselines and Ablation Studies}
Recall that our method, \ours{}, consists of two stages. In the patch feature extraction stage, we introduce VPU with multi-scale inputs. For slide-level aggregation, we propose CrossViT to learn complementary information of multiple scales. We demonstrate the effectiveness of \ours{} by comparing it with alternative strategies in both the patch feature extraction and slide-level aggregation stages.

\subsubsection{Comparison with Alternative Feature Extractors} 
\paragraph{Comparison methods} We propose VPU for patch feature learning with slide-level labels. For fair comparison, we change different feature extractors that do not require patch-level labels and use our proposed CrossViT aggregation for all methods. The comparison feature extractors selected in this study include: 1) pretrained VGG16, ResNet50 on ImageNet \citep{deng2009imagenet}, 2) pretrained feature extractor KimiaNet on histopathology WSI patches, 3) self-supervised contrastive learning with SimCLR \citep{chen2020simple} and DINO  \citep{caron2021emerging}, and 4) weakly-supervised learning, including Positive-negative (PN) learning, which is to train ResNet50 from scratch with slide labels as patch labels. 

\paragraph{Results}
The results of unlearned, self-supervised and weakly-supervised feature extractors are summarized in Table~\ref{tab:03-feature_extractor}. ResNet50 and VGG16 are pretrained on ImageNet \citep{deng2009imagenet}. KimiaNet is another pretrained network on $1000 \times 1000$ histopathology WSI patches, with the backbone of DenseNet121. It has been shown useful in some histopathology WSI but failed with our cytology WSI dataset (loss does not drop), due to the size and content difference between tissue sections and liquid-based cells. Although \citep{chen2022self} claimed that using feature extractor pretrained on real-world datasets will not involve huge drop in histopathology WSI classification performance, we still show that learning-based feature extractor models show superiority over all pretrained feature extractor models in terms of cytology WSIs. Here, weakly-supervised VPU surpasses the best self-supervised learning method by over $2.87\%, 4.26\%, 2.90\%, 0.52\%$ in terms of accuracy, AUC, sensitivity and specificity due to the additional medical background knowledge to guide the learning process. 
\begin{table}[]
\centering
\caption{Comparison on feature extractors with aligned multi-scaled CrossViT. The first section uses pretrained models as feature extractors, the second section uses self-supervised learning to train feature extractors, and the third second section uses weakly-supervised learning to train feature extractors.}
\label{tab:03-feature_extractor}
\resizebox{\columnwidth}{!}{
\begin{tabular}{c|cccc}
\hline \hline
Feature extractor + & \multicolumn{4}{c}{Metrics}              \\ \cline{2-5} 
    Cross-attention     & Acc & AUC & Sensitivity & Specificity \\ \hline\hline
VGG16&
$75.34_{(1.78)}$ &
$75.99_{(1.88)}$&
$82.86_{(2.92)}$&
$72.41_{(2.41)}$\\
ResNet50&
$78.48_{(2.50)}$&
$79.29_{(2.48)}$&
$86.54_{(0.40)}$&
$72.66_{(2.80)}$\\
KimiaNet &-&-&-&-\\
\hline
SimCLR&
$80.82_{(1.40)}$&
$80.53_{(1.45)}$&
$84.62_{(2.15)}$&
$77.78_{(2.61)}$\\
DINO \cite{}&
$81.92_{(1.15)}$&
$81.17_{(1.63)}$&
$88.89_{(3.04)}$&
$77.78_{(2.34)}$\\
\hline
PN & 
$78.56_{(2.49)}$&
$78.24_{(2.58)}$& 
$81.29_{(3.81)}$& 
$75.98_{(4.19)}$\\
VPU &
\bm{$84.79_{(2.13)}$}&
\bm{$85.43_{(2.14)}$} &
\bm{$91.79_{(3.19)}$} &
\bm{$78.30_{(2.97)}$}\\
\hline
\end{tabular}}
\end{table}


\subsubsection{Comparison with Alternative Aggregation Methods}
\paragraph{Comparison methods} We use our VPU features and change different feature aggregation methods to show the superiority of aligned multi-scale CrossViT \citep{chen2021crossvit}. The aggregation methods range from operators like mean/max-pooling \citep{lu2021data}, to GNN-based MIL \citep{li2018graph,zhao2020predicting}, ViT \citep{dosovitskiy2020image} and ViT-based TransMIL \citep{shao2021transmil}. For GNN-MIL \citep{li2018graph,zhao2020predicting}, we regard each patch as a node, patch feature embedding as node features, and connect nodes with cosine feature similarity $>0.9$, then train a graph convolutional network (GCN)  \citep{schlichtkrull2018modeling}. 

\paragraph{Results}
With our VPU features, we compare different aggregation methods in Table~\ref{tab:04-aggregation}. 
\begin{table}[]
\centering
\caption{Comparison of aggregation methods with VPU features. All of the aggregation methods utilize VPU features. multi-scale except for \ours{} are implemented by concatenating features at different scales.}
\label{tab:04-aggregation}
\resizebox{\columnwidth}{!}{
\begin{tabular}{c|c|cccc}
\hline \hline
VPU + &\multirow{2}{*}{Scale}& \multicolumn{4}{c}{Metrics} \\ \cline{3-6} 
Aggregation methods    & \  & Acc & AUC & Sensitivity & Specificity \\ \hline\hline
\multirow{2}{*}{MIL (Max-pooling)} &Single &
$77.67_{(2.60)}$&
$77.94_{(3.46)}$&
$84.39_{(7.56)}$&
$72.86_{(4.79)}$\\
&Multi &
$78.29_{(2.00)}$&
$78.80_{(1.99)}$&
$85.45_{(4.82)}$&
$72.84_{(4.70)}$\\
\multirow{2}{*}{MIL (Mean-pooling)} &Single&
$81.81_{(1.23)}$&
$81.71_{(1.43)}$&
$84.80_{2.46)}$&
$77.94_{(1.91)}$\\
&Multi &
$82.12_{(1.96)}$&
$82.08_{(2.35)}$&
$85.49_{(3.85}$&
\bm{$78.44_{(1.61)}$}\\
\hline
\multirow{2}{*}{GNN-MIL} &Single &
$82.74_{(1.69)}$ &
$83.21_{(1.61)}$ &
$88.95_{(3.47)}$ &
$76.96_{(3.21)}$\\
&Multi &
$83.29_{(1.50)}$ &
$83.75_{(1.55)}$ &
$89.28_{(2.81)}$ &
$77.48_{(2.50)}$\\
\hline
\multirow{2}{*}{ViT} &Single &
$83.36_{(2.04)}$ &
$83.67_{(2.44)}$ &
$88.32_{(2.54)}$ &
$78.01_{(2.44)}$  \\
&Multi &
$82.53_{(1.25)}$&
$82.77_{(1.26)}$&
$87.20_{(2.01)}$&
$76.95_{(2.11)}$\\
\multirow{2}{*}{TransMIL} &Single 
&$83.49_{(1.75)}$ &
$82.31_{(1.12)}$ &
$84.12_{(1.85)}$ &
$78.19_{(2.38)}$   \\
&Multi&
$82.65_{(0.97)}$&
$82.84_{(0.94)}$&
$87.07_{(2.25)}$&
$77.62_{(2.30)}$  \\ \hline
CrossViT&Multi &
\bm{$ 84.79_{(2.13)}$} &
\bm{$85.43_{(2.14)}$} &
\bm{$91.79_{(3.19)}$} &
$78.30_{(2.97)}$   \\ \hline
\end{tabular}}
\end{table}

We divide the aggregation methods into four categories, involving operators like mean/max-pooling, GNN-based MIL, ViT based MIL and multi-path cross-attention based ViT (\ours{}). RNN is not involved as \citep{campanella2019clinical} has shown the limitation of RNN-based aggregation. The input of all methods are VPU features. For each method, we experiment with the single-scale and multi-scale input. We follow \citep{li2021dual} to construct the multi-scale VPU feature by concatenating features at different scales as the input. We observe that our method surpass the best aggregation methods by $1.30\%, 1.76\%, 2.51\%$ in terms of accuracy, AUC and sensitivity. Also, it reaches a comparable result ($78.30\%$) to the best specificity ($78.44\%$). 


\subsubsection{Comparison with Other MIL Baselines for WSI Classification}
\paragraph{Comparison methods}
To demonstrate the effectiveness of the proposed \ours{} on cytological WSI analysis, we compare with state-of-the-art MIL methods for WSIs, including 1) attention-based aggregation ABMIL~\citep{ilse2018attention}, 2) clustering-constrained-attention multiple-instance learning (CLAM)~\citep{lu2021data}, 3) Transformer based correlated MIL (TransMIL)~\citep{shao2021transmil}, and 4) double-tier feature distillation aggregation (DTFT-MIL)~\citep{zhang2022dtfd}. Note that these methods are originally designed for histopathology WSIs analysis using pretrained feature extractors and single-scale inputs. We also choose to include pooling operators such as mean-pooling and max-pooling, following the implementation of \citep{lu2021data}).

Since the preprocessing of most histopathology WSI analysis methods may fail on cytology WSIs (cannot extract any patches, especially for benign WSIs), we use the same patches of our methods when comparing with all baselines except for DSMIL. DSMIL restricts the relationship of magnification between small and large patches ($4\times$). When implementing it, we re-crop $512\times512$ and $128\times128$ patches from large patches with the method provided by the authors (entropy $>5$). When implementing HIPT, we construct two-layer hierarchical features with our $256\times256$ patches and four $128\times128$ patches in each $256\times256$ patch. 

\paragraph{Results}
The classification results of state-of-the-art MIL methods on the cytology WSI dataset are summarized in Table~\ref{tab:02-SOTA}. With the same or more (DSMIL) data, the results suggest that \ours{} brings over $0.68\%, 1.50\%, 3.89\%$ improvement on accuracy, AUC and sensitivity on the WSI classification task. Since we re-crop patches of DSMIL, the average patch number per WSI is $2730$, which is far lager than $100$ used in \ours{}. However, the performance is still $4.59\%, 5.00\%, 7.08\%$ and $1.38\%$ worse than \ours{} in terms of accuracy, AUC, sensitivity and specificity, showing that the result is not sensitive to the number of patches for cytology WSI classification.
\begin{table}[]
\centering
\caption{Comparison with state-of-the-art MIL Methods. The first two sections use pretrained ResNet50 as the feature extractor, the third and fourth sections use self-supervised learning and weakly-supervised to train feature extractors. }
\label{tab:02-SOTA}
\resizebox{\columnwidth}{!}{
\begin{tabular}{c|cccc}
\hline \hline
\multirow{2}{*}{Method} & \multicolumn{4}{c}{Metrics}              \\ \cline{2-5} 
                         & Acc & AUC & Sensitivity & Specificity \\ \hline\hline
MIL (Max-pooling)&$75.96_{(3.25)}$& $75.80_{(3.50)}$&$79.88_{(4.78)}$&$72.02_{(4.37)}$ \\ 
MIL (Mean-pooling) &$76.78_{(1.75)}$&$76.98_{(1.70)}$&$81.89_{(1.84)}$&$71.42_{(2.32)}$ \\
\hline
ABMIL \cite{} &$77.50_{(1.43)}$&$78.32_{(1.21)}$&$85.90_{(1.23)}$& $69.96_{(2.19)}$   \\
CLAM-SB \cite{}&$78.38_{(1.92)}$&$79.28_{(2.47)}$&$87.80_{(2.00)}$&$69.50_{(1.50)}$         \\
TransMIL \cite{}&$78.96_{(2.20)}$&$80.10_{(1.40)}$&$87.80_{(2.30)}$&$71.50_{(0.80)}$\\
DTFD-MIL\cite{}&$78.38_{(2.74)}$&$79.41_{(2.44)}$&$87.90_{(1.62)}$&$70.33_{(3.37)}$         \\
\hline
DSMIL&$80.20_{(3.13)}$&$80.43_{(2.02)}$&$84.71_{(4.76)}$&$76.92_{(4.90)}$\\
HIPT&$84.11_{(1.23)}$&
$83.93_{(1.53)}$&
$86.34_{(2.86)}$&
\bm{$82.50_{(4.74)}$} \\
\hline
\ours{}&\bm{$ 84.79_{(2.13)}$} &\bm{$85.43_{(2.14)}$} &\bm{$91.79_{(3.19)}$} &$78.30_{(2.97)}$  \\ \hline
\end{tabular}}
\end{table}

\subsection{Ablation Study}

\begin{figure}[t]
\centering
\includegraphics[width=\columnwidth]{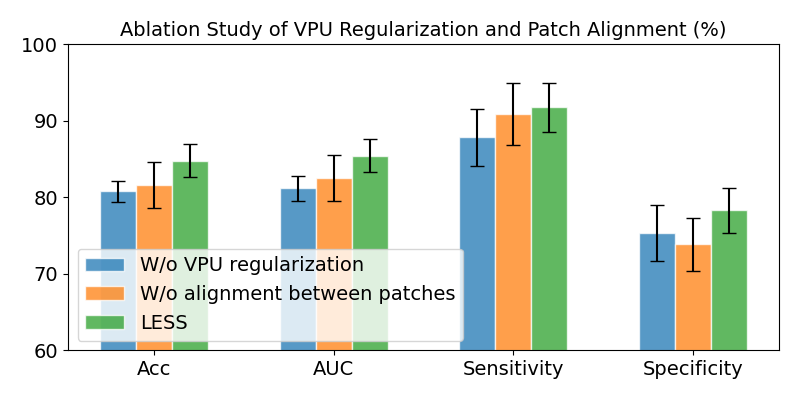}
\vspace{-8mm}
\caption{Ablation Study Results of VPU Regularization and Patch Alignment.} \label{fig:ablation}
\end{figure}

\paragraph{VPU Regularization Term}
We do ablation study experiment on the consistency regularization term in our objective function \eqref{eq:8} of VPU feature extractor. We compare the result (shown in Fig.~\ref{fig:ablation}) of with and without the consistency regularization term to verify its effectiveness in WSI classification.
\paragraph{VPU without Patch Alignment}
As shown in Fig.~\ref{fig:ablation}, when large and small patches do not align with each other, it may bring interference information to the cross-attention module, reducing the performance. When using inaccurate patch labels to weakly-supervise feature extractor learning, it does not bring improvement ($0.08\%$ better on accuracy but $1.03\%$ worse on AUC) compared to directly extracting features with ResNet50.
\paragraph{Single-scale v.s. Multi-scale}
We compare the results of single-scale and multi-scale in Table.~\ref{tab:04-aggregation} as another ablation study. Aggregations apart from attention-based methods perform over $0.5\%$ better in all metrics with concatenated multi-scale features than with single-scale. For attention-based methods, the performance of concatenated feature input is worse than single-scale by at most $0.84\%$ in accuracy, $0.90\%$ in AUC, $1.06\%$ in Specificity. The results prove that early-fusion feature in channel dimension do not necessarily bring improvement to Transformers as described in Sec.~\ref{sec:3.2.2}.

\subsection{Hyperparameter Sensitivity Analysis}
In existing PU learning works, the true label of the testing set is known, and thus the true testing accuracy can be calculated. However, the ratio of benign and malignant patches is agnostic in the real world and will change with the number of randomly selected patches. Therefore, we use a posteriori approach to decide when to stop our VPU model according to the WSI screening result. Additionally, CrossViT has multiple structures based on various embedding token dimensions and numbers of heads, including tiny ($96$ and $192$), small ($192$ and $384$) and base ($384$ and $768$) models  \citep{chen2021crossvit}. We experiment on different CrossViT structures by extracting features of different dimensions.

\begin{figure}[t]
\centering
\subfigure[VPU epoch]{
\begin{minipage}[c]{0.5\columnwidth}
\centering \label{fig:hyper_VPU_epoch}
\includegraphics[width=4.5cm]{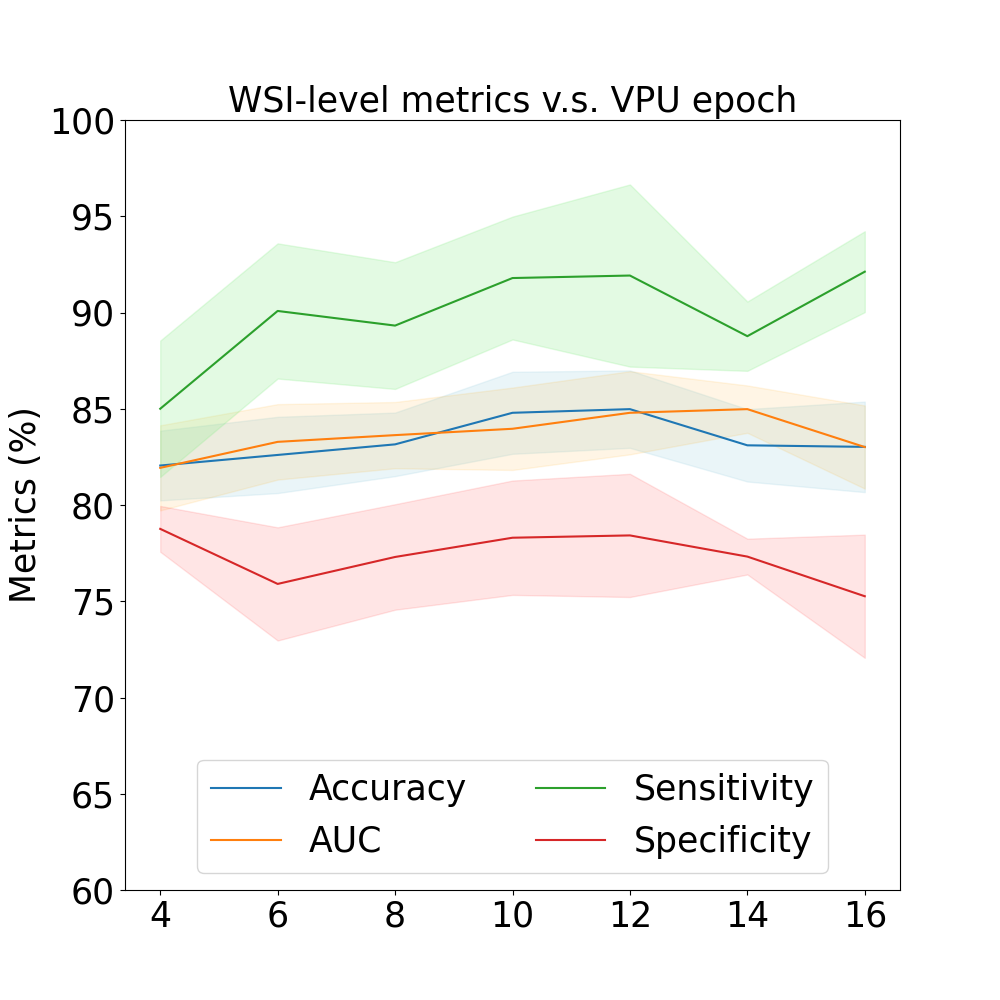}
\end{minipage}%
}%
\subfigure[VPU feature dimension]{
\begin{minipage}[c]{0.5\columnwidth}
\centering \label{fig:hyper_feature_dimension}
\includegraphics[width=4.5cm]{MEDIMA-template/section/fig/hyper_vpu_ep_10_10.png}
\end{minipage}%
}%
\centering
\caption{Hyperparameter sensitivity analysis on VPU epoch and embedding dimension}
\end{figure}

\paragraph{Analysis of VPU Stopping Epoch} 
The number of VPU training epochs is a hyperparameter of \ours{}, which may affect the embedding results. In Fig.~\ref{fig:hyper_VPU_epoch}, the results show that the WSI classification result is not significantly sensitive to the epoch number of VPU, 
meaning it is robust to choose a wide range of stopping epochs for VPU to generate patch feature embedding.
\paragraph{Analysis of VPU Feature Dimension} 
We compare the feature embedding dimensions of VPU in Fig.~\ref{fig:hyper_feature_dimension}. 
With the number of features increasing, the accuracy of WSI screening rises from $82.47\%$ to $84.79\%$ and then decreases to $83.22\%$. Also, higher embedding dimensions require more computational resources and training time, which may not be practical for clinical setups. In our implementation, we set the epoch number to $10$, an earlier point, to save time and computing costs.


\subsection{Interpretation and Visualization}
\begin{figure*}[t!]
\centering
\includegraphics[width=\textwidth]{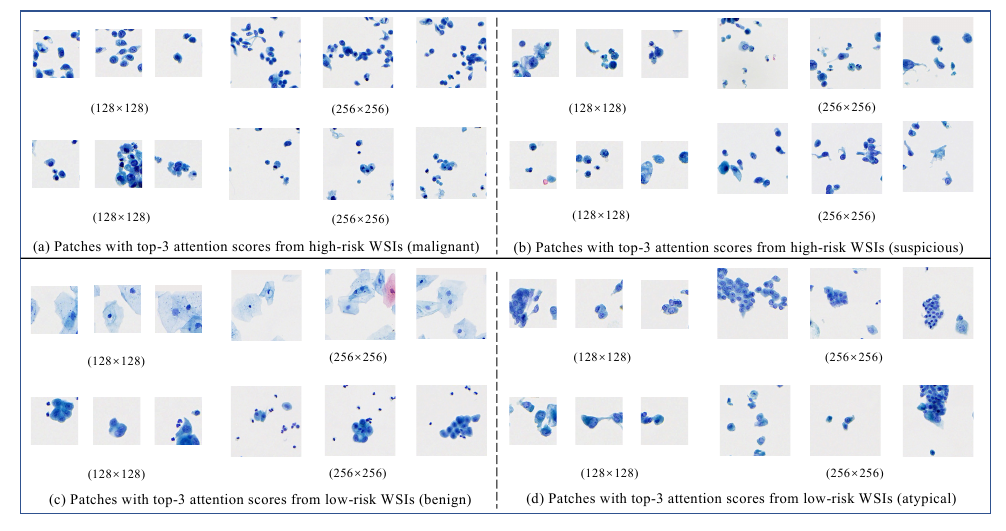}
\caption{$128\times 128$ (left) and $256 \times 256$ (right) patches with top-3 highest attention scores are shown. Each block means a subtype of WSIs containing two examples, with the upper two blocks of high-risk and the lower two of low-risk. The row in the block means patches from the same WSI. The column means patches of the same resolution. } \label{fig:attention_top5}
\end{figure*}
To demonstrate the effectiveness of the cross-attention module, we utilize the attention score of its output class tokens to highlight patches with the highest attention scores. Although we perform binary classification, for better interpretability, we showcase the explainability of our model on four subtypes with attention maps. In Fig.~\ref{fig:attention_top5}, we present two exemplary WSIs per subtype from the testing data. 
Small and big patches are with top-3 highest attention scores are displayed. These patches are expected to contain information that strongly supports clinical judgment. Additionally, large and small patches should not overlap in content, as we aim to extract complementary information from multiple scales.

First, we observe that patches with the highest attention scores at different resolutions from the same slide do not overlap, indicating that \ours{} effectively leverages complementary information from multi-scale inputs. Second, patches with the highest attention scores reflect the critical clinical characteristics of each class, suggesting alignment with the clinical diagnostic criteria.
Specifically, high-grade urothelial carcinoma is evident in 
Fig.~\ref{fig:attention_top5}(a), where numerous tumor cells demonstrate polymorphism, high N/C ratio, prominent nucleoli, irregular nuclei control and hypochromasia. Fig.~\ref{fig:attention_top5}(b) contains cells that are suspicious of high-grade urothelial carcinoma and a few normal intermediate urothelial cells. For suspicious cells, the N/C ratio increases, the chromatin has irregular hypochromasia, and the nuclear membrane exhibits marked irregularities. Fig.~\ref{fig:attention_top5}(c) shows cells of two typical kinds of benign slides, including superficial benign urothelial cells and intermediate or parabasal urothelial cells. The former have frothy and abundant cytoplasm and pale, finely granular chromatin in the nuclei. The later have a similar size and characteristics as superficial cells but have less cytoplasm. Despite the high N/C ratio, they do not have other characteristics of urothelial carcinoma and are present in a neat honeycomb arrangement. Fig.~\ref{fig:attention_top5}(d) shows atypical urothelial cells with a high N/C ratio and an irregular nuclear contour. The absence of hyperchromasia and the presence of degenerated clumped chromatin preclude the diagnosis of suspicious or malignancy.   

\subsection{Results on A Public Dataset}
\begin{table}[]
\centering
\caption{Results $(\%)$ of WSI screening metrics for all baselines and \ours{} on FNAC 2019 dataset. Mean and standard deviation are reported. Results in bold come from \ours{}.}
\label{tab:06-FNAC}
\resizebox{\columnwidth}{!}{
\begin{tabular}{c|c|cccc}
\hline \hline
\multirow{2}{*}{MIL Methods} & \multirow{2}{*}{Scale}&\multicolumn{4}{c}{Metrics}              \\ \cline{3-6} 
                         & & Acc & AUC & Sensitivity & Specificity \\ \hline\hline
\multirow{2}{*}{MIL (Max-pooling)} 
&Single&
$68.49_{(3.03)}$&
$66.39_{(3.24)}$&
\bm{$100.00_{(0.00)}$}&
$62.47_{(2.63)}$\\ 
&Multi&
$74.45_{(3.01)}$&
$72.78_{(3.18)}$&
$98.92_{(2.63)}$&
$67.71_{(2.41)}$\\ 
\multirow{2}{*}{MIL (Mean-pooling)}  
&Single&
$93.75_{(2.34)}$&
$93.33_{(2.50)}$&
\bm{$100.00_{(0.00)}$}&
$89.61_{(3.57)}$\\
&Multi&
$93.96_{(2.34)}$&
$93.56_{(2.59)}$&
\bm{$100.00_{(0.00)}$}&
$89.93_{(3.69)}$\\ 
\hline
\multirow{2}{*}{GNN-MIL}  
&Single& 
$95.63_{(2.04)}$&
$92.36_{(4.09)}$&
\bm{$100.00_{(0.00)}$}&
$94.24_{(2.35)}$\\
&Multi&
$95.98_{(1.77)}$&
$94.97_{(1.56)}$&
$98.75_{(2.80)}$&
$94.97_{(1.84)}$\\\hline
\multirow{2}{*}{ViT} 
&Single& 
$95.90_{(0.97)}$&
$95.85_{(0.98)}$&
$96.23_{(1.61)}$&
$95.66_{(1.42)}$\\ 
&Multi&
$96.33_{(0.76)}$&
$96.30_{(0.79)}$&
$96.33_{(0.95)}$&
$96.36_{(1.23)}$\\ 
\multirow{2}{*}{TransMIL} 
&Single& 
$96.01_{(1.56)}$&
$95.81_{(1.70)}$&
$98.81_{(1.59)}$&
$95.87_{(3.21)}$\\
&Multi& 
$96.48_{(0.86)}$&
$96.30_{(0.93)}$&
$99.16_{(1.49)}$&
$94.46_{(1.87)}$
\\  \hline
HIPT&Multi& 
$96.62_{(1.18)}$&
$96.62_{(1.11)}$&
$96.18_{(2.42)}$&
$97.04_{(0.66)}$\\
\ours{}&Multi&
\bm{$96.88_{(1.03)}$}&
\bm{$96.86_{(0.76)}$}&
$98.95_{(1.34)}$&
\bm{$97.06_{(2.12)}$} \\ \hline
\end{tabular}}
\vspace{-2mm}
\end{table}


We choose to conduct experiments on the FNAC 2019  \citep{saikia2019comparative}, a publicly available dataset containing high-resolution breast cytological images truncated from whole slides. This choice aims to validate the generalizability of \ours{} across different-sized cytological images and cancer types. By testing our model on a diverse set of images from the public dataset, we can evaluate its generalization ability and assess its performance across a broader range of image features. 
The results of \ours{} on the FNAC dataset, compared to baselines models, are shown in Table.~\ref{tab:06-FNAC}. For breast cancer screening, \ours{} shows promising results when compared to the baselines. It achieves an accuracy of $96.88\%$, surpassing the baselines by over $0.26\%$. Moreover, \ours{} outperforms the baselines in terms of AUC, achieving $96.86\%$, which is higher by over $0.24\%$. Additionally, our model exhibits a sensitivity of $98.95\%$ and a specificity of $97.06\%$, which are comparable to the baselines. These results illustrate that our model is highly accurate and specific in detecting abnormalities in a wide range of cytological images, underscoring the potential of our proposed model, \ours{}, to enhance diagnostic accuracy and efficiency in various clinical scenarios.

\section{Discussion}
\label{sec:discussion}
\paragraph{Highlights} In this work, we have developed a deep learning framework to address the challenges of robust patch feature learning and emulating clinical screening across multiple scales. Specifically, we employ a VPU feature encoder to learn patch feature embedding using only slide-level labels. Additionally, our model leverages multi-scale cross-attention mechanism to incorporate comprehensive and complementary information from different scales. One of the primary contributions of our work is the elimination the reliance on fine-grained annotations, which not only save time and labor but also offers generalizability to various cancers screening tasks. The weak supervision of VPU learning model, informed by medical background knowledge, can capture features not pre-defined or recognized, allowing us to refine feature representations based on the underlying true data distribution. This task-alignment supervision in two stages enhances the training process, making it more effective and efficient compared to self-supervised contrastive learning methods or direct utilization of features generated from models pre-trained on real-world datasets.
Furthermore, our approach can capture complementary information from multi-path patches to learn more comprehensive features, leading to more confident decision-making.
\paragraph{Cytological v.s. Histological WSI Analysis}
Most state-of-the-arts MIL methods without reliance of manual annotations are designed for histopathology WSIs. However, when we directly implement those methods to cytological WSIs, the performance is not as good as in histology due to the data property differences. Cytology primarily focuses on assessing the quality and quantity of individual cells, enabling early and rapid cancer or disease screening and diagnosis. In contrast, histopathology involves more detailed tissue and organ analysis. Consequently, reducing magnification during analysis is not a wise choice in cytology. Moreover, cells in cytological WSIs are at a uniformly and randomly distributed, making them more susceptible to artifacts, distortion, and cell overlapping \cite{jiang2023survey}. In contrast, cells in histopathological WSIs provide a more accurate representation of tissue architecture. Therefore, cytology WSI screening task is more challenging to achieve high performance across all metrics. 
\paragraph{Difference from Multi-magnification}
In histopathology WSI analysis, achieving multi-magnification is often done by subsampling the image to create a pyramid-like structure. Each successive level of the pyramid has a lower magnification than the previous one. Lower magnification is used to capture coarse texture information, while higher magnification is employed to capture fine-grained cellular details.However, in cytology WSIs, cells are randomly distributed, lacking significant texture and structural information.Lower magnification images in cytology tend to be blurred and uninformative (see Fig.~\ref{fig:dataset}(a)). Therefore, in our approach, we opt for a multi-scale strategy rather than traditional multi-magnification.
\paragraph{Difference from Previous Multi-scale}
In the original CrossViT  \citep{chen2021crossvit}, patches are extracted with sizes $14 \times 14$ and $16 \times 16$ from $256\times256$ images(randomly cropped to $224\times224$)  at the same magnification level. Large and small patches are not related in any specific way. However, in our preprocessing stage, we take a different approach. We begin by cropping $128 \times 128$ patches with cells in the center, designating them as small patches.Subsequently, we expand these small patches to a size of $256 \times 256$, incorporating the surrounding cells to create large patches. Small patches provide information about individual cells or cell clusters, while large patches offer contextual information. The relationship between our large and small patches is one of inclusion and complementarity. This is why our aligned multi-scaled patches yield better results than randomly selected patches, as shown in Table~\ref{tab:03-feature_extractor}. Additionally, our interpretation (Fig.~\ref{fig:attention_top5}) demonstrates that small and large patches with the highest attention scores are distinct and complementary to each other.
\paragraph{Limitation of PU Learning}
The challenge in PU learning revolves around training a classifier with data that consists of positive and unlabeled examples, which limits its applicability to multi-class datasets. 
In our dataset, WSIs can be catogorized into two risk classes: low-risk and high-risk.However, if a more granular classification is desired, such as categorizing them into benign, atypical, suspicious, and malignant groups, Multi-Positive and Unlabeled learning (MPU) is a possible solution. An approach to address the MPU problem is presented in~\citep{xu2017multi}. The MPU solution involves learning from labeled data originating from multiple positive classes and unlabeled data that can be from the positive classes or an unknown negative class. It does so by learning a discriminant function, denoted as $F: \mathcal{X} \times \mathcal{Z} \rightarrow \mathbb{R}$ over the input and the encoded output pairs $\left(\mathbf{x}_1, \mathbf{z}_1\right), \cdots,\left(\mathbf{x}_n, \mathbf{z}_n\right) \in \mathcal{X} \times \mathcal{Z}$. This method generates a new output space $\mathcal{Z}$ by encoding the original output space $\mathcal{Y}$. However, like most prior work predating VPU \citep{chen2020variational}, it requires estimating the class prior as described in \citep{blanchard2010semi}, which may be challenging to obtain from patches without fine-grained annotations.  Therefore, the design of an MPU method that does not rely on class prior information and can be applied to learn multi-class WSI patch feature representations remains an open question and a promising area for future research.


\section{Conclusion}
This paper introduces a novel label-efficient WSI screening model \ours{}, designed to learn reliable patch features, even from small datasets. It also emulates clinical screening by extracting complementary information from multiple scales. 
The task-alignment property of two stages achieves the best effectiveness and efficiency in the WSI screening task. Furthermore, this methodology can be extended to various applications in WSI screening, offering an effective and efficient clinical solution with substantial value. Notably, our approach eliminates the need for fine-grained annotations, resulting in significant time and manpower savings in the annotation process while achieving superior performance over state-of-the-art baselines.

\section*{Acknowledgments}
This work is supported in part by the Natural Sciences and Engineering Research Council of Canada (NSERC), NVIDIA Hardware Award, Compute Canada and Health Innovation Funding Investment (HIFI) Awards from the University of British Columbia. We thank Dr. SangMook Kim for his meticulous editing, Chun-yin Huang, Ruinan Jin, Yilin Yang for their proofreading.






\bibliographystyle{model2-names.bst}\biboptions{authoryear}
\bibliography{MEDIMA-template/section/reference}

\begin{thebibliography}{73}
\expandafter\ifx\csname natexlab\endcsname\relax\def\natexlab#1{#1}\fi
\providecommand{\url}[1]{\texttt{#1}}
\providecommand{\href}[2]{#2}
\providecommand{\path}[1]{#1}
\providecommand{\DOIprefix}{doi:}
\providecommand{\ArXivprefix}{arXiv:}
\providecommand{\URLprefix}{URL: }
\providecommand{\Pubmedprefix}{pmid:}
\providecommand{\doi}[1]{\href{http://dx.doi.org/#1}{\path{#1}}}
\providecommand{\Pubmed}[1]{\href{pmid:#1}{\path{#1}}}
\providecommand{\bibinfo}[2]{#2}
\ifx\xfnm\relax \def\xfnm[#1]{\unskip,\space#1}\fi
\bibitem[{Awan et~al.(2021)Awan, Benes, Azam, Song, Shaban, Verrill, Tsang,
  Snead, Minhas and Rajpoot}]{awan2021deep}
\bibinfo{author}{Awan, R.}, \bibinfo{author}{Benes, K.}, \bibinfo{author}{Azam,
  A.}, \bibinfo{author}{Song, T.H.}, \bibinfo{author}{Shaban, M.},
  \bibinfo{author}{Verrill, C.}, \bibinfo{author}{Tsang, Y.W.},
  \bibinfo{author}{Snead, D.}, \bibinfo{author}{Minhas, F.},
  \bibinfo{author}{Rajpoot, N.}, \bibinfo{year}{2021}.
\newblock \bibinfo{title}{Deep learning based digital cell profiles for risk
  stratification of urine cytology images}.
\newblock \bibinfo{journal}{Cytometry Part A} \bibinfo{volume}{99},
  \bibinfo{pages}{732--742}.
\bibitem[{Ba et~al.(2016)Ba, Kiros and Hinton}]{ba2016layer}
\bibinfo{author}{Ba, J.L.}, \bibinfo{author}{Kiros, J.R.},
  \bibinfo{author}{Hinton, G.E.}, \bibinfo{year}{2016}.
\newblock \bibinfo{title}{Layer normalization}.
\newblock \bibinfo{journal}{arXiv preprint arXiv:1607.06450} .
\bibitem[{Biewald(2020)}]{wandb}
\bibinfo{author}{Biewald, L.}, \bibinfo{year}{2020}.
\newblock \bibinfo{title}{Experiment tracking with weights and biases}.
\newblock \URLprefix \url{https://www.wandb.com/}. \bibinfo{note}{software
  available from wandb.com}.
\bibitem[{Blanchard et~al.(2010)Blanchard, Lee and Scott}]{blanchard2010semi}
\bibinfo{author}{Blanchard, G.}, \bibinfo{author}{Lee, G.},
  \bibinfo{author}{Scott, C.}, \bibinfo{year}{2010}.
\newblock \bibinfo{title}{Semi-supervised novelty detection}.
\newblock \bibinfo{journal}{The Journal of Machine Learning Research}
  \bibinfo{volume}{11}, \bibinfo{pages}{2973--3009}.
\bibitem[{Campanella et~al.(2019)Campanella, Hanna, Geneslaw, Miraflor, Werneck
  Krauss~Silva, Busam, Brogi, Reuter, Klimstra and
  Fuchs}]{campanella2019clinical}
\bibinfo{author}{Campanella, G.}, \bibinfo{author}{Hanna, M.G.},
  \bibinfo{author}{Geneslaw, L.}, \bibinfo{author}{Miraflor, A.},
  \bibinfo{author}{Werneck Krauss~Silva, V.}, \bibinfo{author}{Busam, K.J.},
  \bibinfo{author}{Brogi, E.}, \bibinfo{author}{Reuter, V.E.},
  \bibinfo{author}{Klimstra, D.S.}, \bibinfo{author}{Fuchs, T.J.},
  \bibinfo{year}{2019}.
\newblock \bibinfo{title}{Clinical-grade computational pathology using weakly
  supervised deep learning on whole slide images}.
\newblock \bibinfo{journal}{Nature medicine} \bibinfo{volume}{25},
  \bibinfo{pages}{1301--1309}.
\bibitem[{Cao et~al.(2021)Cao, Yang, Rong, Li, Xia, You, Lou, Jiang, Du, Meng
  et~al.}]{cao2021novel}
\bibinfo{author}{Cao, L.}, \bibinfo{author}{Yang, J.}, \bibinfo{author}{Rong,
  Z.}, \bibinfo{author}{Li, L.}, \bibinfo{author}{Xia, B.},
  \bibinfo{author}{You, C.}, \bibinfo{author}{Lou, G.}, \bibinfo{author}{Jiang,
  L.}, \bibinfo{author}{Du, C.}, \bibinfo{author}{Meng, H.}, et~al.,
  \bibinfo{year}{2021}.
\newblock \bibinfo{title}{A novel attention-guided convolutional network for
  the detection of abnormal cervical cells in cervical cancer screening}.
\newblock \bibinfo{journal}{Medical image analysis} \bibinfo{volume}{73},
  \bibinfo{pages}{102197}.
\bibitem[{Caron et~al.(2021)Caron, Touvron, Misra, J{\'e}gou, Mairal,
  Bojanowski and Joulin}]{caron2021emerging}
\bibinfo{author}{Caron, M.}, \bibinfo{author}{Touvron, H.},
  \bibinfo{author}{Misra, I.}, \bibinfo{author}{J{\'e}gou, H.},
  \bibinfo{author}{Mairal, J.}, \bibinfo{author}{Bojanowski, P.},
  \bibinfo{author}{Joulin, A.}, \bibinfo{year}{2021}.
\newblock \bibinfo{title}{Emerging properties in self-supervised vision
  transformers}, in: \bibinfo{booktitle}{Proceedings of the IEEE/CVF
  international conference on computer vision}, pp.
  \bibinfo{pages}{9650--9660}.
\bibitem[{Chankong et~al.(2014)Chankong, Theera-Umpon and
  Auephanwiriyakul}]{chankong2014automatic}
\bibinfo{author}{Chankong, T.}, \bibinfo{author}{Theera-Umpon, N.},
  \bibinfo{author}{Auephanwiriyakul, S.}, \bibinfo{year}{2014}.
\newblock \bibinfo{title}{Automatic cervical cell segmentation and
  classification in pap smears}.
\newblock \bibinfo{journal}{Computer methods and programs in biomedicine}
  \bibinfo{volume}{113}, \bibinfo{pages}{539--556}.
\bibitem[{Chen et~al.(2021)Chen, Fan and Panda}]{chen2021crossvit}
\bibinfo{author}{Chen, C.F.R.}, \bibinfo{author}{Fan, Q.},
  \bibinfo{author}{Panda, R.}, \bibinfo{year}{2021}.
\newblock \bibinfo{title}{Crossvit: Cross-attention multi-scale vision
  transformer for image classification}, in: \bibinfo{booktitle}{Proceedings of
  the IEEE/CVF international conference on computer vision}, pp.
  \bibinfo{pages}{357--366}.
\bibitem[{Chen et~al.(2020a)Chen, Liu, Wang, Zhao and Wu}]{chen2020variational}
\bibinfo{author}{Chen, H.}, \bibinfo{author}{Liu, F.}, \bibinfo{author}{Wang,
  Y.}, \bibinfo{author}{Zhao, L.}, \bibinfo{author}{Wu, H.},
  \bibinfo{year}{2020}a.
\newblock \bibinfo{title}{A variational approach for learning from positive and
  unlabeled data}.
\newblock \bibinfo{journal}{Advances in Neural Information Processing Systems}
  \bibinfo{volume}{33}, \bibinfo{pages}{14844--14854}.
\bibitem[{Chen et~al.(2022)Chen, Chen, Li, Chen, Trister, Krishnan and
  Mahmood}]{chen2022scaling}
\bibinfo{author}{Chen, R.J.}, \bibinfo{author}{Chen, C.}, \bibinfo{author}{Li,
  Y.}, \bibinfo{author}{Chen, T.Y.}, \bibinfo{author}{Trister, A.D.},
  \bibinfo{author}{Krishnan, R.G.}, \bibinfo{author}{Mahmood, F.},
  \bibinfo{year}{2022}.
\newblock \bibinfo{title}{Scaling vision transformers to gigapixel images via
  hierarchical self-supervised learning}, in: \bibinfo{booktitle}{Proceedings
  of the IEEE/CVF Conference on Computer Vision and Pattern Recognition}, pp.
  \bibinfo{pages}{16144--16155}.
\bibitem[{Chen and Krishnan(2022)}]{chen2022self}
\bibinfo{author}{Chen, R.J.}, \bibinfo{author}{Krishnan, R.G.},
  \bibinfo{year}{2022}.
\newblock \bibinfo{title}{Self-supervised vision transformers learn visual
  concepts in histopathology}.
\newblock \bibinfo{journal}{arXiv preprint arXiv:2203.00585} .
\bibitem[{Chen et~al.(2020b)Chen, Kornblith, Norouzi and
  Hinton}]{chen2020simple}
\bibinfo{author}{Chen, T.}, \bibinfo{author}{Kornblith, S.},
  \bibinfo{author}{Norouzi, M.}, \bibinfo{author}{Hinton, G.},
  \bibinfo{year}{2020}b.
\newblock \bibinfo{title}{A simple framework for contrastive learning of visual
  representations}, in: \bibinfo{booktitle}{International conference on machine
  learning}, \bibinfo{organization}{PMLR}. pp. \bibinfo{pages}{1597--1607}.
\bibitem[{Chen et~al.(2020c)Chen, Fan, Girshick and He}]{chen2020improved}
\bibinfo{author}{Chen, X.}, \bibinfo{author}{Fan, H.},
  \bibinfo{author}{Girshick, R.}, \bibinfo{author}{He, K.},
  \bibinfo{year}{2020}c.
\newblock \bibinfo{title}{Improved baselines with momentum contrastive
  learning}.
\newblock \bibinfo{journal}{arXiv preprint arXiv:2003.04297} .
\bibitem[{Cheng et~al.(2021)Cheng, Liu, Yu, Rao, Xiao, Han, Zhu, Lv, Li, Cai
  et~al.}]{cheng2021robust}
\bibinfo{author}{Cheng, S.}, \bibinfo{author}{Liu, S.}, \bibinfo{author}{Yu,
  J.}, \bibinfo{author}{Rao, G.}, \bibinfo{author}{Xiao, Y.},
  \bibinfo{author}{Han, W.}, \bibinfo{author}{Zhu, W.}, \bibinfo{author}{Lv,
  X.}, \bibinfo{author}{Li, N.}, \bibinfo{author}{Cai, J.}, et~al.,
  \bibinfo{year}{2021}.
\newblock \bibinfo{title}{Robust whole slide image analysis for cervical cancer
  screening using deep learning}.
\newblock \bibinfo{journal}{Nature communications} \bibinfo{volume}{12},
  \bibinfo{pages}{5639}.
\bibitem[{Coates et~al.(2011)Coates, Ng and Lee}]{coates2011analysis}
\bibinfo{author}{Coates, A.}, \bibinfo{author}{Ng, A.}, \bibinfo{author}{Lee,
  H.}, \bibinfo{year}{2011}.
\newblock \bibinfo{title}{An analysis of single-layer networks in unsupervised
  feature learning}, in: \bibinfo{booktitle}{Proceedings of the fourteenth
  international conference on artificial intelligence and statistics},
  \bibinfo{organization}{JMLR Workshop and Conference Proceedings}. pp.
  \bibinfo{pages}{215--223}.
\bibitem[{Coudray et~al.(2018)Coudray, Ocampo, Sakellaropoulos, Narula,
  Snuderl, Feny{\"o}, Moreira, Razavian and
  Tsirigos}]{coudray2018classification}
\bibinfo{author}{Coudray, N.}, \bibinfo{author}{Ocampo, P.S.},
  \bibinfo{author}{Sakellaropoulos, T.}, \bibinfo{author}{Narula, N.},
  \bibinfo{author}{Snuderl, M.}, \bibinfo{author}{Feny{\"o}, D.},
  \bibinfo{author}{Moreira, A.L.}, \bibinfo{author}{Razavian, N.},
  \bibinfo{author}{Tsirigos, A.}, \bibinfo{year}{2018}.
\newblock \bibinfo{title}{Classification and mutation prediction from
  non--small cell lung cancer histopathology images using deep learning}.
\newblock \bibinfo{journal}{Nature medicine} \bibinfo{volume}{24},
  \bibinfo{pages}{1559--1567}.
\bibitem[{Davey et~al.(2006)Davey, Barratt, Irwig, Chan, Macaskill, Mannes and
  Saville}]{davey2006effect}
\bibinfo{author}{Davey, E.}, \bibinfo{author}{Barratt, A.},
  \bibinfo{author}{Irwig, L.}, \bibinfo{author}{Chan, S.F.},
  \bibinfo{author}{Macaskill, P.}, \bibinfo{author}{Mannes, P.},
  \bibinfo{author}{Saville, A.M.}, \bibinfo{year}{2006}.
\newblock \bibinfo{title}{Effect of study design and quality on unsatisfactory
  rates, cytology classifications, and accuracy in liquid-based versus
  conventional cervical cytology: a systematic review}.
\newblock \bibinfo{journal}{The Lancet} \bibinfo{volume}{367},
  \bibinfo{pages}{122--132}.
\bibitem[{Dehaene et~al.(2020)Dehaene, Camara, Moindrot, de~Lavergne and
  Courtiol}]{dehaene2020selfdehaene2020self}
\bibinfo{author}{Dehaene, O.}, \bibinfo{author}{Camara, A.},
  \bibinfo{author}{Moindrot, O.}, \bibinfo{author}{de~Lavergne, A.},
  \bibinfo{author}{Courtiol, P.}, \bibinfo{year}{2020}.
\newblock \bibinfo{title}{Self-supervision closes the gap between weak and
  strong supervision in histology}.
\newblock \bibinfo{journal}{arXiv preprint arXiv:2012.03583} .
\bibitem[{Deng et~al.(2009)Deng, Dong, Socher, Li, Li and
  Fei-Fei}]{deng2009imagenet}
\bibinfo{author}{Deng, J.}, \bibinfo{author}{Dong, W.},
  \bibinfo{author}{Socher, R.}, \bibinfo{author}{Li, L.J.},
  \bibinfo{author}{Li, K.}, \bibinfo{author}{Fei-Fei, L.},
  \bibinfo{year}{2009}.
\newblock \bibinfo{title}{Imagenet: A large-scale hierarchical image database},
  in: \bibinfo{booktitle}{2009 IEEE conference on computer vision and pattern
  recognition}, \bibinfo{organization}{Ieee}. pp. \bibinfo{pages}{248--255}.
\bibitem[{Devlin et~al.(2018)Devlin, Chang, Lee and Toutanova}]{devlin2018bert}
\bibinfo{author}{Devlin, J.}, \bibinfo{author}{Chang, M.W.},
  \bibinfo{author}{Lee, K.}, \bibinfo{author}{Toutanova, K.},
  \bibinfo{year}{2018}.
\newblock \bibinfo{title}{Bert: Pre-training of deep bidirectional transformers
  for language understanding}.
\newblock \bibinfo{journal}{arXiv preprint arXiv:1810.04805} .
\bibitem[{Dey et~al.(2018)}]{dey2018basic}
\bibinfo{author}{Dey, P.}, et~al., \bibinfo{year}{2018}.
\newblock \bibinfo{title}{Basic and advanced laboratory techniques in
  histopathology and cytology}.
\newblock \bibinfo{publisher}{Springer}.
\bibitem[{Dietterich et~al.(1997)Dietterich, Lathrop and
  Lozano-P{\'e}rez}]{dietterich1997solving}
\bibinfo{author}{Dietterich, T.G.}, \bibinfo{author}{Lathrop, R.H.},
  \bibinfo{author}{Lozano-P{\'e}rez, T.}, \bibinfo{year}{1997}.
\newblock \bibinfo{title}{Solving the multiple instance problem with
  axis-parallel rectangles}.
\newblock \bibinfo{journal}{Artificial intelligence} \bibinfo{volume}{89},
  \bibinfo{pages}{31--71}.
\bibitem[{Donnelly et~al.(2013)Donnelly, Mukherjee, Lyden, Bridge, Lele,
  Wright, McGaughey, Culberson, Horn, Wedel et~al.}]{donnelly2013optimal}
\bibinfo{author}{Donnelly, A.D.}, \bibinfo{author}{Mukherjee, M.S.},
  \bibinfo{author}{Lyden, E.R.}, \bibinfo{author}{Bridge, J.A.},
  \bibinfo{author}{Lele, S.M.}, \bibinfo{author}{Wright, N.},
  \bibinfo{author}{McGaughey, M.F.}, \bibinfo{author}{Culberson, A.M.},
  \bibinfo{author}{Horn, A.J.}, \bibinfo{author}{Wedel, W.R.}, et~al.,
  \bibinfo{year}{2013}.
\newblock \bibinfo{title}{Optimal z-axis scanning parameters for gynecologic
  cytology specimens}.
\newblock \bibinfo{journal}{Journal of pathology informatics}
  \bibinfo{volume}{4}, \bibinfo{pages}{38}.
\bibitem[{Dosovitskiy et~al.(2020)Dosovitskiy, Beyer, Kolesnikov, Weissenborn,
  Zhai, Unterthiner, Dehghani, Minderer, Heigold, Gelly
  et~al.}]{dosovitskiy2020image}
\bibinfo{author}{Dosovitskiy, A.}, \bibinfo{author}{Beyer, L.},
  \bibinfo{author}{Kolesnikov, A.}, \bibinfo{author}{Weissenborn, D.},
  \bibinfo{author}{Zhai, X.}, \bibinfo{author}{Unterthiner, T.},
  \bibinfo{author}{Dehghani, M.}, \bibinfo{author}{Minderer, M.},
  \bibinfo{author}{Heigold, G.}, \bibinfo{author}{Gelly, S.}, et~al.,
  \bibinfo{year}{2020}.
\newblock \bibinfo{title}{An image is worth 16x16 words: Transformers for image
  recognition at scale}.
\newblock \bibinfo{journal}{arXiv preprint arXiv:2010.11929} .
\bibitem[{Du~Plessis et~al.(2014)Du~Plessis, Niu and Sugiyama}]{du2014analysis}
\bibinfo{author}{Du~Plessis, M.C.}, \bibinfo{author}{Niu, G.},
  \bibinfo{author}{Sugiyama, M.}, \bibinfo{year}{2014}.
\newblock \bibinfo{title}{Analysis of learning from positive and unlabeled
  data}.
\newblock \bibinfo{journal}{Advances in neural information processing systems}
  \bibinfo{volume}{27}.
\bibitem[{Dua et~al.(2017)Dua, Graff et~al.}]{dua2017uci}
\bibinfo{author}{Dua, D.}, \bibinfo{author}{Graff, C.}, et~al.,
  \bibinfo{year}{2017}.
\newblock \bibinfo{title}{Uci machine learning repository} .
\bibitem[{Garud et~al.(2017)Garud, Karri, Sheet, Chatterjee, Mahadevappa, Ray,
  Ghosh and Maity}]{garud2017high}
\bibinfo{author}{Garud, H.}, \bibinfo{author}{Karri, S.P.K.},
  \bibinfo{author}{Sheet, D.}, \bibinfo{author}{Chatterjee, J.},
  \bibinfo{author}{Mahadevappa, M.}, \bibinfo{author}{Ray, A.K.},
  \bibinfo{author}{Ghosh, A.}, \bibinfo{author}{Maity, A.K.},
  \bibinfo{year}{2017}.
\newblock \bibinfo{title}{High-magnification multi-views based classification
  of breast fine needle aspiration cytology cell samples using fusion of
  decisions from deep convolutional networks}, in:
  \bibinfo{booktitle}{Proceedings of the IEEE conference on computer vision and
  pattern recognition workshops}, pp. \bibinfo{pages}{76--81}.
\bibitem[{He et~al.(2016)He, Zhang, Ren and Sun}]{he2016deep}
\bibinfo{author}{He, K.}, \bibinfo{author}{Zhang, X.}, \bibinfo{author}{Ren,
  S.}, \bibinfo{author}{Sun, J.}, \bibinfo{year}{2016}.
\newblock \bibinfo{title}{Deep residual learning for image recognition}, in:
  \bibinfo{booktitle}{Proceedings of the IEEE conference on computer vision and
  pattern recognition}, pp. \bibinfo{pages}{770--778}.
\bibitem[{Hendrycks and Gimpel(2016)}]{hendrycks2016gaussian}
\bibinfo{author}{Hendrycks, D.}, \bibinfo{author}{Gimpel, K.},
  \bibinfo{year}{2016}.
\newblock \bibinfo{title}{Gaussian error linear units (gelus)}.
\newblock \bibinfo{journal}{arXiv preprint arXiv:1606.08415} .
\bibitem[{Hou et~al.(2017)Hou, Chaib-Draa, Li and Zhao}]{hou2017generative}
\bibinfo{author}{Hou, M.}, \bibinfo{author}{Chaib-Draa, B.},
  \bibinfo{author}{Li, C.}, \bibinfo{author}{Zhao, Q.}, \bibinfo{year}{2017}.
\newblock \bibinfo{title}{Generative adversarial positive-unlabelled learning}.
\newblock \bibinfo{journal}{arXiv preprint arXiv:1711.08054} .
\bibitem[{Ilse et~al.(2018)Ilse, Tomczak and Welling}]{ilse2018attention}
\bibinfo{author}{Ilse, M.}, \bibinfo{author}{Tomczak, J.},
  \bibinfo{author}{Welling, M.}, \bibinfo{year}{2018}.
\newblock \bibinfo{title}{Attention-based deep multiple instance learning}, in:
  \bibinfo{booktitle}{International conference on machine learning},
  \bibinfo{organization}{PMLR}. pp. \bibinfo{pages}{2127--2136}.
\bibitem[{Jiang et~al.(2022)Jiang, Zhou, Lin, Chan, Liu and
  Chen}]{jiang2022deep}
\bibinfo{author}{Jiang, H.}, \bibinfo{author}{Zhou, Y.}, \bibinfo{author}{Lin,
  Y.}, \bibinfo{author}{Chan, R.C.}, \bibinfo{author}{Liu, J.},
  \bibinfo{author}{Chen, H.}, \bibinfo{year}{2022}.
\newblock \bibinfo{title}{Deep learning for computational cytology: A survey}.
\newblock \bibinfo{journal}{Medical Image Analysis} , \bibinfo{pages}{102691}.
\bibitem[{Jiang et~al.(2023)Jiang, Li, Shen, Chen, Wang, Chen, Feng and
  Liu}]{jiang2023survey}
\bibinfo{author}{Jiang, P.}, \bibinfo{author}{Li, X.}, \bibinfo{author}{Shen,
  H.}, \bibinfo{author}{Chen, Y.}, \bibinfo{author}{Wang, L.},
  \bibinfo{author}{Chen, H.}, \bibinfo{author}{Feng, J.}, \bibinfo{author}{Liu,
  J.}, \bibinfo{year}{2023}.
\newblock \bibinfo{title}{A survey on deep learning-based cervical cytology
  screening: from cell identification to whole slide image analysis} .
\bibitem[{Kiryo et~al.(2017)Kiryo, Niu, Du~Plessis and
  Sugiyama}]{kiryo2017positive}
\bibinfo{author}{Kiryo, R.}, \bibinfo{author}{Niu, G.},
  \bibinfo{author}{Du~Plessis, M.C.}, \bibinfo{author}{Sugiyama, M.},
  \bibinfo{year}{2017}.
\newblock \bibinfo{title}{Positive-unlabeled learning with non-negative risk
  estimator}.
\newblock \bibinfo{journal}{Advances in neural information processing systems}
  \bibinfo{volume}{30}.
\bibitem[{Kitchener et~al.(2006)Kitchener, Castle and
  Cox}]{kitchener2006achievements}
\bibinfo{author}{Kitchener, H.C.}, \bibinfo{author}{Castle, P.E.},
  \bibinfo{author}{Cox, J.T.}, \bibinfo{year}{2006}.
\newblock \bibinfo{title}{Achievements and limitations of cervical cytology
  screening}.
\newblock \bibinfo{journal}{Vaccine} \bibinfo{volume}{24},
  \bibinfo{pages}{S63--S70}.
\bibitem[{Krizhevsky et~al.(2009)Krizhevsky, Hinton
  et~al.}]{krizhevsky2009learning}
\bibinfo{author}{Krizhevsky, A.}, \bibinfo{author}{Hinton, G.}, et~al.,
  \bibinfo{year}{2009}.
\newblock \bibinfo{title}{Learning multiple layers of features from tiny
  images} .
\bibitem[{Kurian et~al.(2022)Kurian, Lehan, Verghese, Dharamshi, Meena, Li,
  Liu, Gillet, Rane, Grigoriadis et~al.}]{kurian2022deep}
\bibinfo{author}{Kurian, N.C.}, \bibinfo{author}{Lehan, A.},
  \bibinfo{author}{Verghese, G.}, \bibinfo{author}{Dharamshi, N.},
  \bibinfo{author}{Meena, S.}, \bibinfo{author}{Li, M.}, \bibinfo{author}{Liu,
  F.}, \bibinfo{author}{Gillet, C.}, \bibinfo{author}{Rane, S.},
  \bibinfo{author}{Grigoriadis, A.}, et~al., \bibinfo{year}{2022}.
\newblock \bibinfo{title}{Deep multi-scale u-net architecture and label-noise
  robust training strategies for histopathological image segmentation}, in:
  \bibinfo{booktitle}{2022 IEEE 22nd International Conference on Bioinformatics
  and Bioengineering (BIBE)}, \bibinfo{organization}{IEEE}. pp.
  \bibinfo{pages}{91--96}.
\bibitem[{Li et~al.(2021a)Li, Li and Eliceiri}]{li2021dual}
\bibinfo{author}{Li, B.}, \bibinfo{author}{Li, Y.}, \bibinfo{author}{Eliceiri,
  K.W.}, \bibinfo{year}{2021}a.
\newblock \bibinfo{title}{Dual-stream multiple instance learning network for
  whole slide image classification with self-supervised contrastive learning},
  in: \bibinfo{booktitle}{Proceedings of the IEEE/CVF conference on computer
  vision and pattern recognition}, pp. \bibinfo{pages}{14318--14328}.
\bibitem[{Li et~al.(2021b)Li, Chen, Huang, Yang, Hu, Duan, Metaxas, Li and
  Zhang}]{li2021hybrid}
\bibinfo{author}{Li, J.}, \bibinfo{author}{Chen, W.}, \bibinfo{author}{Huang,
  X.}, \bibinfo{author}{Yang, S.}, \bibinfo{author}{Hu, Z.},
  \bibinfo{author}{Duan, Q.}, \bibinfo{author}{Metaxas, D.N.},
  \bibinfo{author}{Li, H.}, \bibinfo{author}{Zhang, S.}, \bibinfo{year}{2021}b.
\newblock \bibinfo{title}{Hybrid supervision learning for pathology whole slide
  image classification}, in: \bibinfo{booktitle}{Medical Image Computing and
  Computer Assisted Intervention--MICCAI 2021: 24th International Conference,
  Strasbourg, France, September 27--October 1, 2021, Proceedings, Part VIII},
  \bibinfo{organization}{Springer}. pp. \bibinfo{pages}{309--318}.
\bibitem[{Li et~al.(2018)Li, Yao, Zhu, Li and Huang}]{li2018graph}
\bibinfo{author}{Li, R.}, \bibinfo{author}{Yao, J.}, \bibinfo{author}{Zhu, X.},
  \bibinfo{author}{Li, Y.}, \bibinfo{author}{Huang, J.}, \bibinfo{year}{2018}.
\newblock \bibinfo{title}{Graph cnn for survival analysis on whole slide
  pathological images}, in: \bibinfo{booktitle}{International Conference on
  Medical Image Computing and Computer-Assisted Intervention},
  \bibinfo{organization}{Springer}. pp. \bibinfo{pages}{174--182}.
\bibitem[{Lotan and Roehrborn(2003)}]{lotan2003sensitivity}
\bibinfo{author}{Lotan, Y.}, \bibinfo{author}{Roehrborn, C.G.},
  \bibinfo{year}{2003}.
\newblock \bibinfo{title}{Sensitivity and specificity of commonly available
  bladder tumor markers versus cytology: results of a comprehensive literature
  review and meta-analyses}.
\newblock \bibinfo{journal}{Urology} \bibinfo{volume}{61},
  \bibinfo{pages}{109--118}.
\bibitem[{Lu et~al.(2021)Lu, Williamson, Chen, Chen, Barbieri and
  Mahmood}]{lu2021data}
\bibinfo{author}{Lu, M.Y.}, \bibinfo{author}{Williamson, D.F.},
  \bibinfo{author}{Chen, T.Y.}, \bibinfo{author}{Chen, R.J.},
  \bibinfo{author}{Barbieri, M.}, \bibinfo{author}{Mahmood, F.},
  \bibinfo{year}{2021}.
\newblock \bibinfo{title}{Data-efficient and weakly supervised computational
  pathology on whole-slide images}.
\newblock \bibinfo{journal}{Nature biomedical engineering} \bibinfo{volume}{5},
  \bibinfo{pages}{555--570}.
\bibitem[{Maron and Lozano-P{\'e}rez(1997)}]{maron1997framework}
\bibinfo{author}{Maron, O.}, \bibinfo{author}{Lozano-P{\'e}rez, T.},
  \bibinfo{year}{1997}.
\newblock \bibinfo{title}{A framework for multiple-instance learning}.
\newblock \bibinfo{journal}{Advances in neural information processing systems}
  \bibinfo{volume}{10}.
\bibitem[{Mehrotra et~al.(2011)Mehrotra, Mishra, Singh and
  Singh}]{mehrotra2011efficacy}
\bibinfo{author}{Mehrotra, R.}, \bibinfo{author}{Mishra, S.},
  \bibinfo{author}{Singh, M.}, \bibinfo{author}{Singh, M.},
  \bibinfo{year}{2011}.
\newblock \bibinfo{title}{The efficacy of oral brush biopsy with
  computer-assisted analysis in identifying precancerous and cancerous
  lesions}.
\newblock \bibinfo{journal}{Head \& neck oncology} \bibinfo{volume}{3},
  \bibinfo{pages}{1--8}.
\bibitem[{Morrison and DeNicola(1993)}]{morrison1993advantages}
\bibinfo{author}{Morrison, W.}, \bibinfo{author}{DeNicola, D.},
  \bibinfo{year}{1993}.
\newblock \bibinfo{title}{Advantages and disadvantages of cytology and
  histopathology for the diagnosis of cancer.}, in:
  \bibinfo{booktitle}{Seminars in veterinary medicine and surgery (small
  animal)}, pp. \bibinfo{pages}{222--227}.
\bibitem[{Nojima et~al.(2021)Nojima, Terayama, Shimoura, Hijiki, Nonomura,
  Morii, Okuno and Fujita}]{nojima2021deep}
\bibinfo{author}{Nojima, S.}, \bibinfo{author}{Terayama, K.},
  \bibinfo{author}{Shimoura, S.}, \bibinfo{author}{Hijiki, S.},
  \bibinfo{author}{Nonomura, N.}, \bibinfo{author}{Morii, E.},
  \bibinfo{author}{Okuno, Y.}, \bibinfo{author}{Fujita, K.},
  \bibinfo{year}{2021}.
\newblock \bibinfo{title}{A deep learning system to diagnose the malignant
  potential of urothelial carcinoma cells in cytology specimens}.
\newblock \bibinfo{journal}{Cancer Cytopathology} \bibinfo{volume}{129},
  \bibinfo{pages}{984--995}.
\bibitem[{Plissiti et~al.(2009)Plissiti, Tripoliti, Charchanti, Krikoni and
  Fotiadis}]{plissiti2009automated}
\bibinfo{author}{Plissiti, M.}, \bibinfo{author}{Tripoliti, E.},
  \bibinfo{author}{Charchanti, A.}, \bibinfo{author}{Krikoni, O.},
  \bibinfo{author}{Fotiadis, D.}, \bibinfo{year}{2009}.
\newblock \bibinfo{title}{Automated detection of cell nuclei in pap stained
  cervical smear images using fuzzy clustering}, in: \bibinfo{booktitle}{4th
  European Conference of the International Federation for Medical and
  Biological Engineering: ECIFMBE 2008 23--27 November 2008 Antwerp, Belgium},
  \bibinfo{organization}{Springer}. pp. \bibinfo{pages}{637--641}.
\bibitem[{Saikia et~al.(2019)Saikia, Bora, Mahanta and
  Das}]{saikia2019comparative}
\bibinfo{author}{Saikia, A.R.}, \bibinfo{author}{Bora, K.},
  \bibinfo{author}{Mahanta, L.B.}, \bibinfo{author}{Das, A.K.},
  \bibinfo{year}{2019}.
\newblock \bibinfo{title}{Comparative assessment of cnn architectures for
  classification of breast fnac images}.
\newblock \bibinfo{journal}{Tissue and Cell} \bibinfo{volume}{57},
  \bibinfo{pages}{8--14}.
\bibitem[{Sanghvi et~al.(2019)Sanghvi, Allen, Callenberg and
  Pantanowitz}]{sanghvi2019performance}
\bibinfo{author}{Sanghvi, A.B.}, \bibinfo{author}{Allen, E.Z.},
  \bibinfo{author}{Callenberg, K.M.}, \bibinfo{author}{Pantanowitz, L.},
  \bibinfo{year}{2019}.
\newblock \bibinfo{title}{Performance of an artificial intelligence algorithm
  for reporting urine cytopathology}.
\newblock \bibinfo{journal}{Cancer Cytopathology} \bibinfo{volume}{127},
  \bibinfo{pages}{658--666}.
\bibitem[{Schlichtkrull et~al.(2018)Schlichtkrull, Kipf, Bloem, Van Den~Berg,
  Titov and Welling}]{schlichtkrull2018modeling}
\bibinfo{author}{Schlichtkrull, M.}, \bibinfo{author}{Kipf, T.N.},
  \bibinfo{author}{Bloem, P.}, \bibinfo{author}{Van Den~Berg, R.},
  \bibinfo{author}{Titov, I.}, \bibinfo{author}{Welling, M.},
  \bibinfo{year}{2018}.
\newblock \bibinfo{title}{Modeling relational data with graph convolutional
  networks}, in: \bibinfo{booktitle}{The Semantic Web: 15th International
  Conference, ESWC 2018, Heraklion, Crete, Greece, June 3--7, 2018, Proceedings
  15}, \bibinfo{organization}{Springer}. pp. \bibinfo{pages}{593--607}.
\bibitem[{Shao et~al.(2021)Shao, Bian, Chen, Wang, Zhang, Ji
  et~al.}]{shao2021transmil}
\bibinfo{author}{Shao, Z.}, \bibinfo{author}{Bian, H.}, \bibinfo{author}{Chen,
  Y.}, \bibinfo{author}{Wang, Y.}, \bibinfo{author}{Zhang, J.},
  \bibinfo{author}{Ji, X.}, et~al., \bibinfo{year}{2021}.
\newblock \bibinfo{title}{Transmil: Transformer based correlated multiple
  instance learning for whole slide image classification}.
\newblock \bibinfo{journal}{Advances in neural information processing systems}
  \bibinfo{volume}{34}, \bibinfo{pages}{2136--2147}.
\bibitem[{Sun et~al.(2021)Sun, Hutchinson, Tomaszewicz, Caporelli, Meng,
  McCauley, Fischer, Cosar and Cornejo}]{sun2021diagnostic}
\bibinfo{author}{Sun, T.}, \bibinfo{author}{Hutchinson, L.},
  \bibinfo{author}{Tomaszewicz, K.}, \bibinfo{author}{Caporelli, M.L.},
  \bibinfo{author}{Meng, X.}, \bibinfo{author}{McCauley, K.},
  \bibinfo{author}{Fischer, A.H.}, \bibinfo{author}{Cosar, E.F.},
  \bibinfo{author}{Cornejo, K.M.}, \bibinfo{year}{2021}.
\newblock \bibinfo{title}{Diagnostic value of a comprehensive, urothelial
  carcinoma--specific next-generation sequencing panel in urine cytology and
  bladder tumor specimens}.
\newblock \bibinfo{journal}{Cancer Cytopathology} \bibinfo{volume}{129},
  \bibinfo{pages}{537--547}.
\bibitem[{Teramoto et~al.(2017)Teramoto, Tsukamoto, Kiriyama and
  Fujita}]{teramoto2017automated}
\bibinfo{author}{Teramoto, A.}, \bibinfo{author}{Tsukamoto, T.},
  \bibinfo{author}{Kiriyama, Y.}, \bibinfo{author}{Fujita, H.},
  \bibinfo{year}{2017}.
\newblock \bibinfo{title}{Automated classification of lung cancer types from
  cytological images using deep convolutional neural networks}.
\newblock \bibinfo{journal}{BioMed research international}
  \bibinfo{volume}{2017}.
\bibitem[{Tolkach et~al.(2020)Tolkach, Dohmg{\"o}rgen, Toma and
  Kristiansen}]{tolkach2020high}
\bibinfo{author}{Tolkach, Y.}, \bibinfo{author}{Dohmg{\"o}rgen, T.},
  \bibinfo{author}{Toma, M.}, \bibinfo{author}{Kristiansen, G.},
  \bibinfo{year}{2020}.
\newblock \bibinfo{title}{High-accuracy prostate cancer pathology using deep
  learning}.
\newblock \bibinfo{journal}{Nature Machine Intelligence} \bibinfo{volume}{2},
  \bibinfo{pages}{411--418}.
\bibitem[{Vaswani et~al.(2017)Vaswani, Shazeer, Parmar, Uszkoreit, Jones,
  Gomez, Kaiser and Polosukhin}]{vaswani2017attention}
\bibinfo{author}{Vaswani, A.}, \bibinfo{author}{Shazeer, N.},
  \bibinfo{author}{Parmar, N.}, \bibinfo{author}{Uszkoreit, J.},
  \bibinfo{author}{Jones, L.}, \bibinfo{author}{Gomez, A.N.},
  \bibinfo{author}{Kaiser, {\L}.}, \bibinfo{author}{Polosukhin, I.},
  \bibinfo{year}{2017}.
\newblock \bibinfo{title}{Attention is all you need}.
\newblock \bibinfo{journal}{Advances in neural information processing systems}
  \bibinfo{volume}{30}.
\bibitem[{Wang et~al.(2023)Wang, Muzakky, Lee, Lin and
  Chao}]{wang2023annotation}
\bibinfo{author}{Wang, C.W.}, \bibinfo{author}{Muzakky, H.},
  \bibinfo{author}{Lee, Y.C.}, \bibinfo{author}{Lin, Y.J.},
  \bibinfo{author}{Chao, T.K.}, \bibinfo{year}{2023}.
\newblock \bibinfo{title}{Annotation-free deep learning-based prediction of
  thyroid molecular cancer biomarker braf (v600e) from cytological slides}.
\newblock \bibinfo{journal}{International Journal of Molecular Sciences}
  \bibinfo{volume}{24}, \bibinfo{pages}{2521}.
\bibitem[{Xiao et~al.(2017)Xiao, Rasul and Vollgraf}]{xiao2017fashion}
\bibinfo{author}{Xiao, H.}, \bibinfo{author}{Rasul, K.},
  \bibinfo{author}{Vollgraf, R.}, \bibinfo{year}{2017}.
\newblock \bibinfo{title}{Fashion-mnist: a novel image dataset for benchmarking
  machine learning algorithms}.
\newblock \bibinfo{journal}{arXiv preprint arXiv:1708.07747} .
\bibitem[{Xu et~al.(2017)Xu, Xu, Xu and Tao}]{xu2017multi}
\bibinfo{author}{Xu, Y.}, \bibinfo{author}{Xu, C.}, \bibinfo{author}{Xu, C.},
  \bibinfo{author}{Tao, D.}, \bibinfo{year}{2017}.
\newblock \bibinfo{title}{Multi-positive and unlabeled learning.}, in:
  \bibinfo{booktitle}{IJCAI}, pp. \bibinfo{pages}{3182--3188}.
\bibitem[{Yao et~al.(2020)Yao, Zhu, Jonnagaddala, Hawkins and
  Huang}]{yao2020whole}
\bibinfo{author}{Yao, J.}, \bibinfo{author}{Zhu, X.},
  \bibinfo{author}{Jonnagaddala, J.}, \bibinfo{author}{Hawkins, N.},
  \bibinfo{author}{Huang, J.}, \bibinfo{year}{2020}.
\newblock \bibinfo{title}{Whole slide images based cancer survival prediction
  using attention guided deep multiple instance learning networks}.
\newblock \bibinfo{journal}{Medical Image Analysis} \bibinfo{volume}{65},
  \bibinfo{pages}{101789}.
\bibitem[{Yu et~al.(2023)Yu, Ma, Fu, Chen, Lai, Zhuo and Xu}]{yu2023local}
\bibinfo{author}{Yu, J.}, \bibinfo{author}{Ma, T.}, \bibinfo{author}{Fu, Y.},
  \bibinfo{author}{Chen, H.}, \bibinfo{author}{Lai, M.}, \bibinfo{author}{Zhuo,
  C.}, \bibinfo{author}{Xu, Y.}, \bibinfo{year}{2023}.
\newblock \bibinfo{title}{Local-to-global spatial learning for whole-slide
  image representation and classification}.
\newblock \bibinfo{journal}{Computerized Medical Imaging and Graphics} ,
  \bibinfo{pages}{102230}.
\bibitem[{Yu et~al.(2022)Yu, Ghosh, Liu, Deible and
  Batmanghelich}]{yu2022anatomy}
\bibinfo{author}{Yu, K.}, \bibinfo{author}{Ghosh, S.}, \bibinfo{author}{Liu,
  Z.}, \bibinfo{author}{Deible, C.}, \bibinfo{author}{Batmanghelich, K.},
  \bibinfo{year}{2022}.
\newblock \bibinfo{title}{Anatomy-guided weakly-supervised abnormality
  localization in chest x-rays}, in: \bibinfo{booktitle}{Medical Image
  Computing and Computer Assisted Intervention--MICCAI 2022: 25th International
  Conference, Singapore, September 18--22, 2022, Proceedings, Part V},
  \bibinfo{organization}{Springer}. pp. \bibinfo{pages}{658--668}.
\bibitem[{Yu et~al.(2020)Yu, Wang, Xiao, Cao, Yue, Liu, Wang, Xu and
  Lei}]{yu2020multi}
\bibinfo{author}{Yu, S.}, \bibinfo{author}{Wang, S.}, \bibinfo{author}{Xiao,
  X.}, \bibinfo{author}{Cao, J.}, \bibinfo{author}{Yue, G.},
  \bibinfo{author}{Liu, D.}, \bibinfo{author}{Wang, T.}, \bibinfo{author}{Xu,
  Y.}, \bibinfo{author}{Lei, B.}, \bibinfo{year}{2020}.
\newblock \bibinfo{title}{Multi-scale enhanced graph convolutional network for
  early mild cognitive impairment detection}, in: \bibinfo{booktitle}{Medical
  Image Computing and Computer Assisted Intervention--MICCAI 2020: 23rd
  International Conference, Lima, Peru, October 4--8, 2020, Proceedings, Part
  VII 23}, \bibinfo{organization}{Springer}. pp. \bibinfo{pages}{228--237}.
\bibitem[{Zerouaoui and Idri(2022)}]{zerouaoui2022deep}
\bibinfo{author}{Zerouaoui, H.}, \bibinfo{author}{Idri, A.},
  \bibinfo{year}{2022}.
\newblock \bibinfo{title}{Deep hybrid architectures for binary classification
  of medical breast cancer images}.
\newblock \bibinfo{journal}{Biomedical Signal Processing and Control}
  \bibinfo{volume}{71}, \bibinfo{pages}{103226}.
\bibitem[{Zerouaoui et~al.(2022)Zerouaoui, Idri and
  Idri}]{zerouaoui2022classifying}
\bibinfo{author}{Zerouaoui, H.}, \bibinfo{author}{Idri, A.},
  \bibinfo{author}{Idri, A.}, \bibinfo{year}{2022}.
\newblock \bibinfo{title}{Classifying breast cytological images using deep
  learning architectures.}, in: \bibinfo{booktitle}{HEALTHINF}, pp.
  \bibinfo{pages}{557--564}.
\bibitem[{Zhang et~al.(2019)Zhang, Ren, Liu, Yang and Gong}]{zhang2019positive}
\bibinfo{author}{Zhang, C.}, \bibinfo{author}{Ren, D.}, \bibinfo{author}{Liu,
  T.}, \bibinfo{author}{Yang, J.}, \bibinfo{author}{Gong, C.},
  \bibinfo{year}{2019}.
\newblock \bibinfo{title}{Positive and unlabeled learning with label
  disambiguation.}, in: \bibinfo{booktitle}{IJCAI}, pp.
  \bibinfo{pages}{4250--4256}.
\bibitem[{Zhang et~al.(2017)Zhang, Cisse, Dauphin and
  Lopez-Paz}]{zhang2017mixup}
\bibinfo{author}{Zhang, H.}, \bibinfo{author}{Cisse, M.},
  \bibinfo{author}{Dauphin, Y.N.}, \bibinfo{author}{Lopez-Paz, D.},
  \bibinfo{year}{2017}.
\newblock \bibinfo{title}{mixup: Beyond empirical risk minimization}.
\newblock \bibinfo{journal}{arXiv preprint arXiv:1710.09412} .
\bibitem[{Zhang et~al.(2022a)Zhang, Meng, Zhao, Qiao, Yang, Coupland and
  Zheng}]{zhang2022dtfd}
\bibinfo{author}{Zhang, H.}, \bibinfo{author}{Meng, Y.}, \bibinfo{author}{Zhao,
  Y.}, \bibinfo{author}{Qiao, Y.}, \bibinfo{author}{Yang, X.},
  \bibinfo{author}{Coupland, S.E.}, \bibinfo{author}{Zheng, Y.},
  \bibinfo{year}{2022}a.
\newblock \bibinfo{title}{Dtfd-mil: Double-tier feature distillation multiple
  instance learning for histopathology whole slide image classification}, in:
  \bibinfo{booktitle}{Proceedings of the IEEE/CVF Conference on Computer Vision
  and Pattern Recognition}, pp. \bibinfo{pages}{18802--18812}.
\bibitem[{Zhang et~al.(2022b)Zhang, Cao, Wang, Sun, Fan, Wang and
  Zhang}]{zhang2022whole}
\bibinfo{author}{Zhang, X.}, \bibinfo{author}{Cao, M.}, \bibinfo{author}{Wang,
  S.}, \bibinfo{author}{Sun, J.}, \bibinfo{author}{Fan, X.},
  \bibinfo{author}{Wang, Q.}, \bibinfo{author}{Zhang, L.},
  \bibinfo{year}{2022}b.
\newblock \bibinfo{title}{Whole slide cervical cancer screening using graph
  attention network and supervised contrastive learning}, in:
  \bibinfo{booktitle}{Medical Image Computing and Computer Assisted
  Intervention--MICCAI 2022: 25th International Conference, Singapore,
  September 18--22, 2022, Proceedings, Part II},
  \bibinfo{organization}{Springer}. pp. \bibinfo{pages}{202--211}.
\bibitem[{Zhao et~al.(2020)Zhao, Yang, Fang, Liu, Zhou, Zhang, Sun, Yang,
  Menze, Fan et~al.}]{zhao2020predicting}
\bibinfo{author}{Zhao, Y.}, \bibinfo{author}{Yang, F.}, \bibinfo{author}{Fang,
  Y.}, \bibinfo{author}{Liu, H.}, \bibinfo{author}{Zhou, N.},
  \bibinfo{author}{Zhang, J.}, \bibinfo{author}{Sun, J.},
  \bibinfo{author}{Yang, S.}, \bibinfo{author}{Menze, B.},
  \bibinfo{author}{Fan, X.}, et~al., \bibinfo{year}{2020}.
\newblock \bibinfo{title}{Predicting lymph node metastasis using
  histopathological images based on multiple instance learning with deep graph
  convolution}, in: \bibinfo{booktitle}{Proceedings of the IEEE/CVF Conference
  on Computer Vision and Pattern Recognition}, pp. \bibinfo{pages}{4837--4846}.
\bibitem[{Zhao et~al.(2021)Zhao, Pang, Liu and Ye}]{zhao2021positive}
\bibinfo{author}{Zhao, Z.}, \bibinfo{author}{Pang, F.}, \bibinfo{author}{Liu,
  Z.}, \bibinfo{author}{Ye, C.}, \bibinfo{year}{2021}.
\newblock \bibinfo{title}{Positive-unlabeled learning for cell detection in
  histopathology images with incomplete annotations}, in:
  \bibinfo{booktitle}{Medical Image Computing and Computer Assisted
  Intervention--MICCAI 2021: 24th International Conference, Strasbourg, France,
  September 27--October 1, 2021, Proceedings, Part VIII 24},
  \bibinfo{organization}{Springer}. pp. \bibinfo{pages}{509--518}.
\bibitem[{Zhu et~al.(2017)Zhu, Yao, Zhu and Huang}]{zhu2017wsisa}
\bibinfo{author}{Zhu, X.}, \bibinfo{author}{Yao, J.}, \bibinfo{author}{Zhu,
  F.}, \bibinfo{author}{Huang, J.}, \bibinfo{year}{2017}.
\newblock \bibinfo{title}{Wsisa: Making survival prediction from whole slide
  histopathological images}, in: \bibinfo{booktitle}{Proceedings of the IEEE
  conference on computer vision and pattern recognition}, pp.
  \bibinfo{pages}{7234--7242}.
\bibitem[{Zuluaga et~al.(2011)Zuluaga, Hush, Delgado~Leyton, Hoyos and
  Orkisz}]{zuluaga2011learning}
\bibinfo{author}{Zuluaga, M.A.}, \bibinfo{author}{Hush, D.},
  \bibinfo{author}{Delgado~Leyton, E.J.}, \bibinfo{author}{Hoyos, M.H.},
  \bibinfo{author}{Orkisz, M.}, \bibinfo{year}{2011}.
\newblock \bibinfo{title}{Learning from only positive and unlabeled data to
  detect lesions in vascular ct images}, in: \bibinfo{booktitle}{Medical Image
  Computing and Computer-Assisted Intervention--MICCAI 2011: 14th International
  Conference, Toronto, Canada, September 18-22, 2011, Proceedings, Part III
  14}, \bibinfo{organization}{Springer}. pp. \bibinfo{pages}{9--16}.

\end{thebibliography}



\end{document}